# Land use change in agricultural systems: an integrated ecological-social simulation model of farmer decisions and cropping system performance based on a cellular automata approach

Ferraro, Diego O.[1]*; Pessah, Sebastián[1]; Blanco, Daniela[2]; and Castro, Rodrigo[23]

[1] Universidad de Buenos Aires. Consejo Nacional de Investigaciones Científicas y Técnicas. Instituto de Investigaciones Fisiológicas y Ecológicas Vinculadas a la Agricultura (IFEVA-CONICET). Facultad de Agronomía, Cátedra de Cerealicultura. Av. San Martín 4453, Buenos Aires 1417DSE, Argentina

[2] Universidad de Buenos Aires. Facultad de Ciencias Exactas y Naturales. Departamento de Computación

[3] Consejo Nacional de Investigaciones Científicas y Técnicas. Instituto de Ciencias de la Computación (ICC-CONICET)

*Corresponding Author:

Telephone: 54-11-4524-8053 (ext. 49)

E-mail: ferraro@agro.edu.ar

Fax: 54-11-4524-8053 (ext.33)

**Abstract**

Agricultural systems changes are driven by population growth and intensification of technological inputs. These forces shape the land-use and cover change (LUCC) dynamics representing a complex landscape transformation process. In order to understand and represent the LUCC process we developed a spatially explicit agent-based model in the form of a Cellular Automata: the AgroDEVS model. The model was designed for predicting LUCC dynamics along with their associated economic and environmental changes. AgroDEVS is structured using behavioral rules and functions representing a) crop yields, b) weather conditions, c) economic profit, d) farmer's preferences, e) technology level adoption and f) natural resources consumption based on embodied energy accounting. Using data from a typical location of the Pampa region (Argentina) for the 1988-2015 period, simulation exercises showed that the economic goals were achieved, on average, each 6 out of 10 years, but the environmental thresholds were only achieved in 1.9 out of 10 years. In a set of 50-years simulations, LUCC patterns quickly converge towards the most profitable crop sequences, with no noticeable tradeoff between the economic and environmental conditions.

**Software availability section**

Name of software

AgroDEVS

Developer:

Daniela Blanco, Rodrigo Castro, Diego O. Ferraro

Contact address:

Discrete Event Simulation Lab. Computer Science Dept., School of Exact Sciences, University of Buenos Aires Intendente Guiraldes 2160, C1428EGA, Ciudad Autonoma de Buenos Aires Argentina. dblanco@dc.uba.ar. rcastro@dc.uba.ar +54-11-4576-3390 ext.717

Year first available:

2016

Hardware required:

Standard general-purpose computer running Linux

Software required:

CD++ general purpose discrete-event simulator (http://cell-devs.sce.carleton.ca/mediawiki/index.php/Main_Page).

PHP

MySQL

Availability and cost:

Contact the authors

Program language:

Model developed under the Cell-DEVS language and web interface developed with PHP and MySQL. Base simulator written in C++ by its original authors.

Access form:

Remote web access can be arranged.

**Introduction**

In agroecosystems, land-use and cover change (LUCC) is driven by simultaneous responses to economic opportunities, institutional factors and environmental constraints (Lambin et al., 2001). Different methodological approaches are used for assessing LUCC by capturing the dynamic process influenced by complex interactions between socio-economic drivers and biophysical conditions. For example, the spatially explicit LUCC models assess location suitability for different land uses and allocate changes to grid cells based on suitability maps (Verburg et al., 2004a). Another modelling approach has been developed for capturing the behavior of the real actors of land-use change: individuals and/or institutions ("agents") become the objects of analysis, modelling and simulation, paying explicit attention to interactions between these agents of change (Castella and Verburg, 2007). In this modelling approach, farmer responses to environmental, economic and sociological constraints for deciding on LUCC are crucial for assessing the agricultural system sustainability. Such models are known as agent-based models (ABM)[1]

The ABM approach provides a flexible paradigm for studying emergent patterns in complex systems (Cook, 2009). Particularly, in agricultural systems the ABMs are capable of capturing the individual agent behavior in response to several constraints scenarios (e.g. climate, socio-economical). In addition, it is possible to develop ABMs into spatially explicit frameworks for exploring the LUCC process over time. Examples may include spatial interaction models, cellular automata (CA) and dynamic system models (Evans and Kelley, 2004). However, when ABM are used to cope with complex systems the development of models often trends to get forced to the ABM paradigm even when certain dynamics are not necessarily well expressed in the ABM realm. In this context, the Discrete Event Systems Specification (DEVS) modelling and simulation framework (Zeigler et al., 2000) offers a universal paradigm for modelling hybrid models (continuous, discrete time, discrete event) that are able to scale up through the interconnection of models of very different nature, including the ABM approach. In the case of environmental systems there exists a considerable modelling experience based on the DEVS formalism (Filippi et al., 2010). Notably, when ABM needs to be combined with spatially explicit dynamics, the Cell-DEVS formalism provides means to attain this goal within the DEVS framework (see (Wainer, 2006) for experiences of Cell-DEVS applied to environmental dynamics). DEVS enforces the formal specifications and strict separation between model specification, abstract simulator algorithm, and experimental framework. This greatly facilitates the modelling and simulation software development efforts, as progress can be made in the model aspects, simulation aspects, or experimentation aspects in an independent fashion, preserving the composability of said technologies. For instance, in this work (Zapatero et al., 2011) the authors developed a DEVS-based software solution where GIS technology, Cell-DEVS models, DEVS distributed simulation and Google Earth visualization were orchestrated to build a fire spread simulation application relying on pre existing components).

---

[1] We use ABM also as an equivalent for IBM (Individual-Based Models) or MAS (Multiagent Systems) see Railsback, S.F., Grimm, V., 2011. Agent-based and individual-based modeling: a practical introduction. Princeton university press.

   Although the ABM approach is widely used to explain land use change and future policies impacts, the development of LUCC simulation models coupled with environment impact assessment is still incipient in agricultural systems (Kremmydas et al., 2018). The first generation of ABM was related to agricultural economics (Balmann, 1997), followed by numbers of studies for simulating the individual farm's performance and their spatial interactions (Berger, 2001; Happe et al., 2008; Schreinemachers et al., 2007). Lately, it is recognized that the incorporation of material and energy flows during land-use conversion are increasingly needed to explore socio-economic dynamics and land-use change (Lee et al., 2008). Thus, the ABM includes other aspects such as policy implications (Happe et al., 2006), dynamic models of environmental processes (Schreinemachers and Berger, 2011), or organizational simulation, market simulation (Bonabeau, 2002). In the studied area (the Pampa region, Argentina) where climate, technological innovations, and socio-economic contexts affect agricultural production, the ABM approach for LUCC modelling has not been frequently addressed (Groeneveld et al., 2017). An exception to this rule is the model of Bert et al. (2011) (the PAMPAS model) for gaining understanding about both structural and land use changes in the Pampas. Although we use a similar modelling approach as the PAMPAS model regarding the agent's behavior, PAMPAS does not include any modelling effort for assessing the status level of the natural resources involved due to the LUCC simulation results.

In this work, we developed an ABM model called AgroDEVS, implemented with the Cell-DEVS cellular automata-oriented formalism that relies on the Discrete Event Systems Specification (DEVS) modelling and simulation framework. The goal of AgroDEVS was to integrate into a single model, the main driving forces (i.e. climate, agronomic management, and farmer decisions) for explaining LUCC as well as the environmental and economic consequences of these changes. This integration is developed into a decision-making tool that could be extended and improved to specific policy contexts. AgroDEVS aims to simulate LUCC as well as economic profit and fossil energy use (as a proxy of the environmental condition), both at the individual (agent) and collective (landscape) scales.

**Materials and Methods**

**Region under study and cropping system description**

AgroDEVS was applied to simulate LUCC dynamics in Pergamino, a typical location in the Rolling Pampas (Figure 1), the most productive subregion of the Pampas where most of the annual cropping is concentrated (Hall et al., 1992). The Pampa is a fertile plain originally covered by grasslands, which during the 1900s and 2000s was transformed into an agricultural land mosaic by grazing and farming activities (Soriano et al., 1991). The predominant soils are typical Arguidols and the annual rainfall averaged 950 mm (Moscatelli et al., 1980). AgroDEVS simulates LUCC using the most frequent crop types in the Pampa region (Manuel-Navarrete et al., 2009): (1) the wheat/soybean double-cropping (W/S); (2) maize cropping (M), and spring soybean cropping (S). In this paper, the terms "crop"; "crop type" and "land use" (i.e. cropping systems) were used interchangeably as well as "farmer" and "agent". The agronomic decisions (i.e. genotype selection, fertilizer management, pest control, sowing date, and soil type) were representative of the most frequent situation for each of the cropping systems.

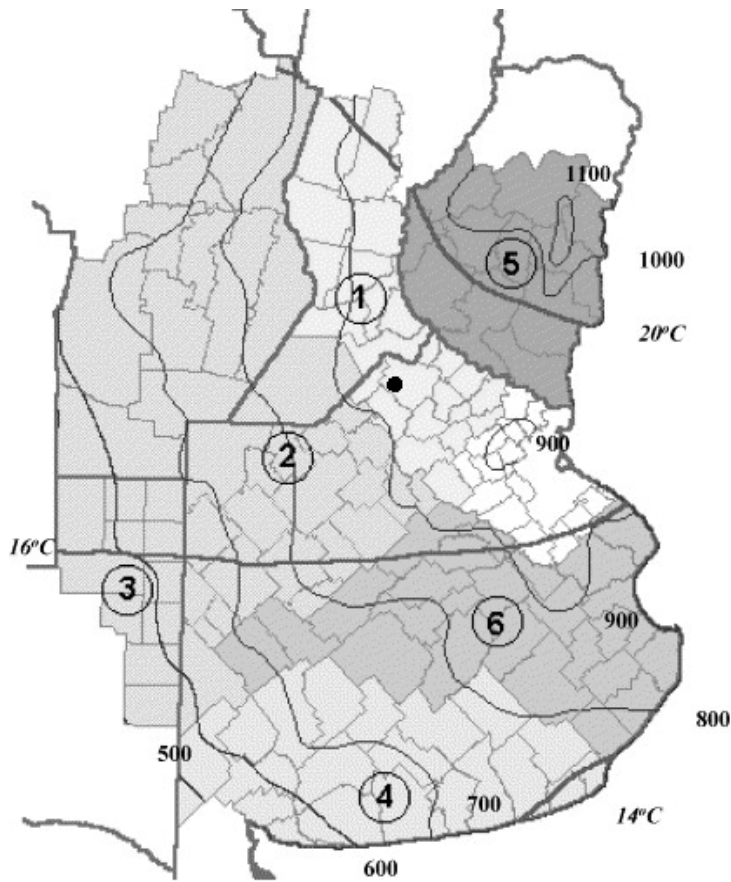

Figure 1: Location of the study site (Pergamino, Buenos Aires) within the Pampa region. (1) rolling pampas, (2) subhumid central pampas, (3) semiarid central pampas, (4) southern pampas, (5) Mesopotamian pampas, and (6) flooding pampas. Thin isolines are isohyets (mm per year); thick isolines are mean annual temperature (°C). Adapted from Viglizzo et al. (2004).

**Cellular Automata-based modelling and simulation approach**

For the modelling activity we followed a scenario-based analysis, exploring agricultural mosaic dynamics. Then we adopted an agent–based modelling (ABM) approach, by identifying both landscape and agent-specific descriptors as parameters (fixed) or attributes (variable). We defined parameters as any fixed condition for describing the behavior or the condition of a model element. On the other hand, the attributes represent the system changes during the model run period. Lastly, we adopted a formal model-based simulation framework to specify mathematically both the parameters and attributes due to different behaviors. For this purpose, we encoded AgroDEVS using the Cell-DEVS formalism. Cell-DEVS is an extension for Cellular Automata of the more generic Discrete EVent System Specification (DEVS) formal modelling and simulation framework. On the one hand, the DEVS formalism permits to express and combine any kind of dynamical system (continuous, discrete event, discrete time) in a mathematical form that is independent of any programming language. On the other hand, Cell-DEVS provides the modeler with a meta-language tailored to facilitate expressing systems where the spatial arrangement of "cells", and their behavior, play a salient role. By undertaking the DEVS-based approach, AgroDEVS becomes readily linkable with other DEVS models developed by others for different domains (e.g. climate, biology, sociology, and economy), potentially using heterogeneous techniques (e.g. differential equations, equilibrium models, optimization models, stochastic processes). In Figure 2, we summarize these concepts. Atomic DEVS are the smallest units of behavior. They can be interconnected modularly through input and output ports to compose hierarchies of more complex systems called Coupled DEVS (Figure 2b). Cell-DEVS features an automatic composition of Atomic DEVS models in the form of an N-dimensional lattice. Each cell gets interconnected only to other cells belonging to a neighborhood shape defined by the modeler (Figure 2a). Cell-DEVS features a rule-based compact language to model the behavior of each cell in relation to its neighborhood, influencing each other.

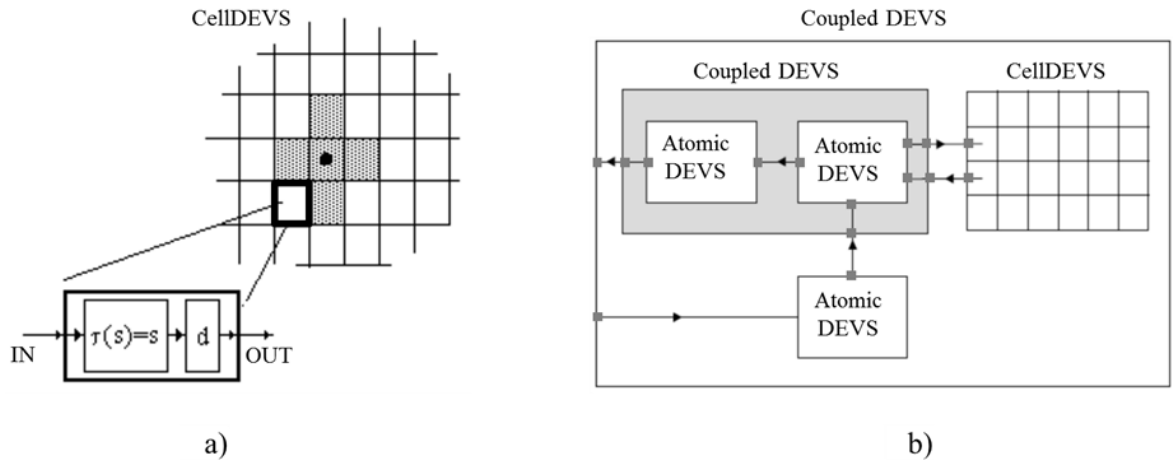

Figure 2. Modular and hierarchical composition of systems with DEVS and Cell-DEVS. a) Cellular Automata oriented Cell-DEVS with a Von Neumann neighborhood (cross-like greyed cells). b) Composition of DEVS and Cell-DEVS models.

Behavioral rules are used in the Cell-DEVS language to define the change of attributes that are local to each cell. These variables can express properties for that unit of space (e.g. in the 2-dimensional case it can be an agricultural plot). When the modeler uses such variables to express also the attributes of "agents" located at cells, then the ABM approach can be merged into the system. In the case of AgroDEVS each agent (namely, a farmer) remains fixed to his or her plot, and rules are used to express changes both for the physical environment and the for farmer, which eventually affect each other. In this work, we adopted the CD++ simulation software toolkit, which is capable of interpreting Cell-DEVS and DEVS models and of simulating them. CD++ implements a generic DEVS "abstract simulator", consisting of a standardized algorithmic recipe that specifies how to simulate any DEVS model independently of any programming language of choice. We believe this generic, reusable and extensible DEVS-based approach provides AgroDEVS with very desirable scalability and sustainability features. The full model represents the collective (i.e. landscape-level) function that emerges from the aggregation of all farmers' outcomes. It also depends on exogenous variables (e.g., climate, crop (output) prices, and production costs) as well as endogenous variables (e.g. the farmer's technological level, the outcomes of neighboring farmers, and each farmer's performance history). The model proved particularly well suited to reproduce empirical situations where (a) there are changes in the relative production/output prices between potential land uses, (b) there is a specific climate regime that impacts on crop yields, or (c) there are varying aspiration levels for the farmers.

In Appendix E we provide more technical details about the model specification approach, its modeling language, rule specification, simulation framework and experimentation environment.

**Model validation and sensitivity analysis**

We submitted AgroDEVS to a systematic verification process (Wilensky and Rand, 2007). Firstly, a code walk-through (Stern, 2003) was performed to review the model formulation and to ensure that all design concepts and specifications are correctly reflected in the code. In addition, AgroDEVS was run with very few farmers in the grid (9–16) and results were examined closely (e.g., following dynamics of specific farmers, inspecting the outliers). The intrinsic complexities and uncertainties on both the magnitude and the nature of driving forces and land use, lead to expand the scope of the straightforward evaluation between simulated and observed patterns in the model validation phase (Bert et al., 2014; Le et al., 2012; Nguyen and de Kok, 2007). Our model development process seeks to validate the ABM through a process of continuous adaptation using feedback from the stakeholders (Ligtenberg et al., 2010). Specifically, AgroDEVS development entailed a continual discussion process with stakeholders (data not shown) from the study area in order to review both the rules and the assumptions of the model that initially came from the literature review. This approach directly engages stakeholders in model development by embedding it within the social process of policy development (Moss, 2008). This changes the validation problem into an advantage: the agreement of participants or stakeholders may be an indicator of the validity of a simulation model (Troitzsch, 2004). In AgroDEVS, the evaluation of the simulation is guided by the expectations, anticipations, and experience of the community that uses it for practical purposes (Ahrweiler and Gilbert, 2005), and this supports the view that the very meaning of validity is dependent on the purpose of the simulation models under examination (Küppers and Lenhard, 2005). Moreover, it is possible to develop a model that fits the data perfectly with a model structure that does not capture any real processes (Cooley and Solano, 2011). Due to the complex nature of ABMs (e.g. nonlinear responses to parameter) a broad single-parameter SA (Appendix A) was performed (Railsback and Grimm, 2015). This type of analysis assesses the effect of each parameter over a wide range of values, as opposed to the traditional local analysis, in which each model input is varied by a standard small amount (Ligmann-Zielinska et al., 2020; Railsback and Grimm, 2015). The goodness-of-fit of each simulated LUCC was calculated using 1) v = RMSE/Observed mean % Total area; 2) PM (probability of a match) = #matches / (#matches + #mismatches); and 3) the index of observed fit (IOF) = (2 x PM) - 1. The number of matches (#matches) is calculated by counting the set of ordered pairs of observations that match the predicted ordered pairs of a simulation (Thorngate and Edmonds, 2013)

## Results

**Model description**

The model description follows the standard protocol "ODD" (Overview, Design concepts, and Details) to standardize the published descriptions of individual-based and agent-based models (Grimm et al., 2010). We described in the following subsections 1) the system variables and 2) the process overview to emphasize the main message of the model outputs. The rest of the ODD protocol items (the initialization conditions, the submodels description, the input data, the modelling approach, and the design concepts) can be found in the Appendixes B to F, respectively.

*System variables*

AgroDEVS maps farmers onto a regular grid in order to initialize the simulations. The model consists of two entities: 1) the **landscape** and 2) the **agents** that operate on the landscape. Each entity has its own set of fixed parameters and attributes that evolve throughout simulation cycles. The landscape parameters are a) the number of agents and b) the owner/tenant agent ratio and the landscape attribute is the overall outcome from the integration of all individual agent attributes results within the simulated landscape. The agent parameters are a) the land rental price (RP) for the tenants, and b) its location on the grid. The attributes for each agent are a) the technological level (TL), b) the crop type allocation (or land use, LU), c) the economic profit (P), d) the renewability level (RL) of the emergy consumption (see Appendix C *Renewability level calculations* section for a emergy concept explanation), e) the aspiration level (AL), f) the environmental threshold (ET), and g) the weather growing condition (WGC).

*Process overview*

AgroDEVS simulations advance with an annual time step, representing a single cropping cycle (CC) (Figure 3). Crop yields under different WGC were previously simulated (Appendix B, Table B1) using the DSSAT model (Jones et al., 2003). We used local management data for defining resource level in each TL (e.g. fertilization regime, genotype) as well as five contrasting historical weather records for defining the different WGC levels.

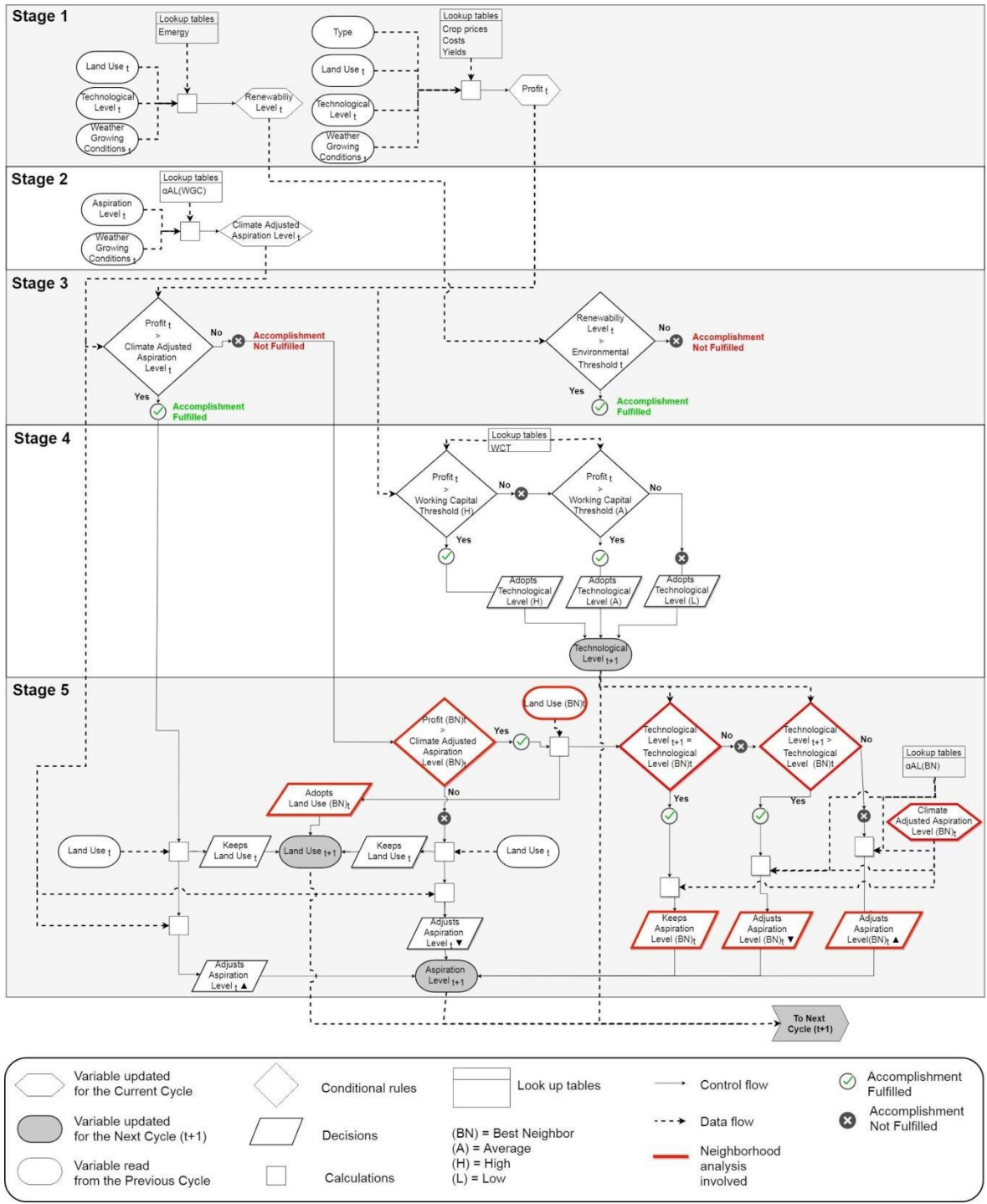

Figure 3 AgroDEVS event flow diagram for each cropping cycle (CC) for a single agent. The stages and their actions are encoded in the form of Cell-DEVS rules. Ovals and hexagons show agent's attributes and attributes calculated during the CC, respectively. Square boxes denote calculations performed at each stage (for more details of each step see Appendix C). Rectangles represent goal fulfillments, diamonds are conditional rules, and parallelograms are an agent's decisions. Solid and dotted lines denote control flow and data flow, respectively. Red borders highlight when the agent performs a neighborhood analysis, and grey background indicates that the calculated value shall be used in the next cropping cycle.

At the start of the AgroDEVS simulation process (Figure 3, Stage 1), each initial agent's configuration ($LU_t$) is exposed to a climate-related growing condition level WGC that range from very favorable to very unfavorable for high crop yield achievement. Then, based on 1) the previously simulated crop yields, 2) the crop price, 3) the production cost of each TL, 4) the land rental price and 5) the crop type allocation into the farm household, AgroDEVS calculates the farmer's $P_t$: the profit P for the current time t, or similarly, for the current cropping cycle CC. At this first step, AgroDEVS also calculates $RL_t$ (see Appendix C for RL calculation details) as a measure of the renewable energy consumption of the cropping cycle. At the next stage, the initial agent's aspiration level ($AL_t$) is adjusted by means of the current $WGC_t$ level resulting in a new climate-adjusted aspiration level ($CAL_t$) (Figure 3, Stage 2). Then, AgroDEVS uses $P_t$ and $RL_t$ values calculated in Stage 1 for assessing the fulfillment of both the environmental (ET) and economic (AL) goals (Figure 3, Stage 3) of each agent. An agent will keep its crop type allocation once the $P_t$ value is greater than or equal to its CAL at each CC ($CAL_t$), while non-fulfillment of environmental threshold (ET) does not alter its crop type allocation decision. AgroDEVS assesses whether the farmer can upgrade or downgrade its TL for the next CC, according to their economic performance in the current CC (Figure 3, Stage 4). For this adjustment, AgroDEVS defines a set of different working capital thresholds (WCT) in order to access different TL values. The WCT fulfillment is assessed using $P_t$. If $P_t$ is lower than the respective WCT, the farmer must lower its TL or vice versa. Eventually, agents can remain at the lowest TL, regardless of its $P_t$ value (i.e. no agent is forced out of business). During this stage, AgroDEVS also adjusts the farmer aspiration level (AL) for the next CC. This setting is based on a) the farmer's perception of the WGC level of the next CC, and b) the agent's failure or success at achieving the AL in the previous CC, respectively. Values of 1) crop prices, 2) adjustment factor of the aspiration level due to WGD, 3) adjustment factor of the aspiration level due to TL, and 4) working capital threshold for TL are shown in the

Appendix Tables B4 to B7, respectively).

**Simulation results for Pergamino 1988-2015**

*LUCC patterns*

The simulated LUCC patterns replicated the overall trend towards soybean-dominated landscapes observed in the region since the mid-1990s. The ordinal pattern analyses (OPA) showed a similar, relatively high probability of a match (PM) for maize, wheat/soybean, and soybean, evidencing the model's capability to predict ordinal (higher/lower) values of crop type cover. However, the accuracy in predicting the magnitude of these changes was lower (Figure 4). Based on the RMSE method, the goodness-of-fit of the wheat/soybean simulated cover was the lowest among the three-crop type, resulting in a significant overestimation of this cropping area. The model underestimated simulated soybean cover. In addition, the v values (i.e. the RMSE relative to the observed mean) was remarkably higher for wheat/soybean than in both maize and soybean.

*Profit, Renewability, and Technological Levels*

Profit (P) and renewability level (RL) values for the Pergamino simulation correlated without any significant trend throughout the study period (Figure 5). Remarkably, in only four years of the simulated period, the simulated landscape exhibited higher RL values than the ecological threshold (ET > 50 % RL). Inter-agent variability in the simulated landscape depicted a range of decisions (in terms of crop type allocation) that resulted in a great variability in average P values during the simulation (Figure 6).

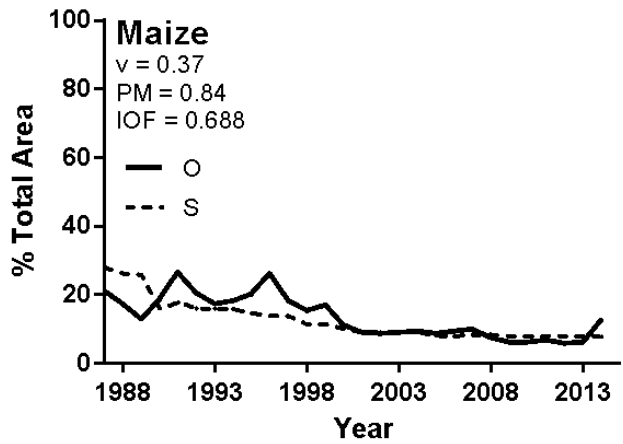

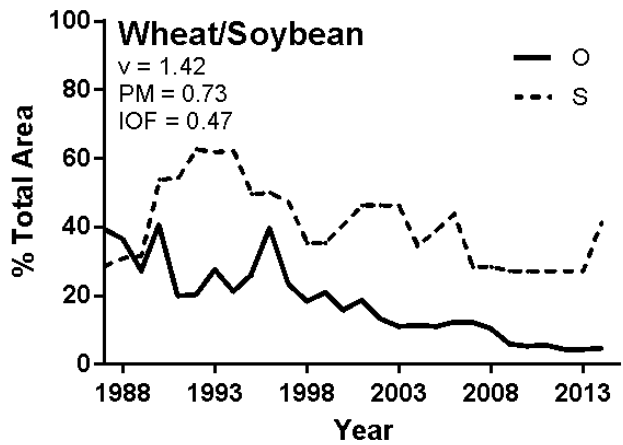

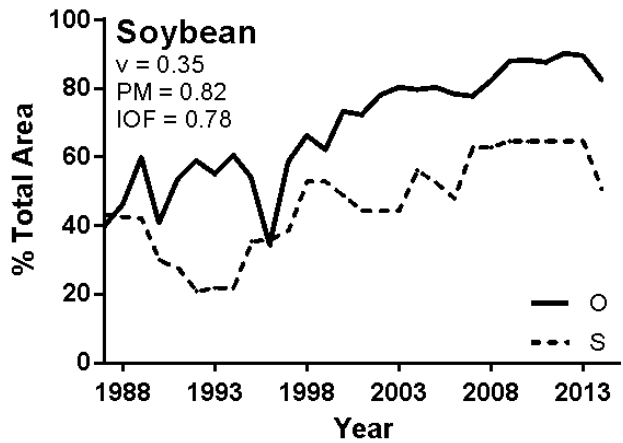

Figure 4. Observed (O) and simulated (S) land use cover, expressed as % of Total Area, for the Pergamino 1988-2014 simulation. The goodness-of-fit of each simulated LU change pattern are 1) v = RMSE/Observed mean % Total area; 2) PM (probability of a match) = #matches / (#matches +

#mismatches); and 3) the index of observed fit (IOF) = (2 x PM) - 1.

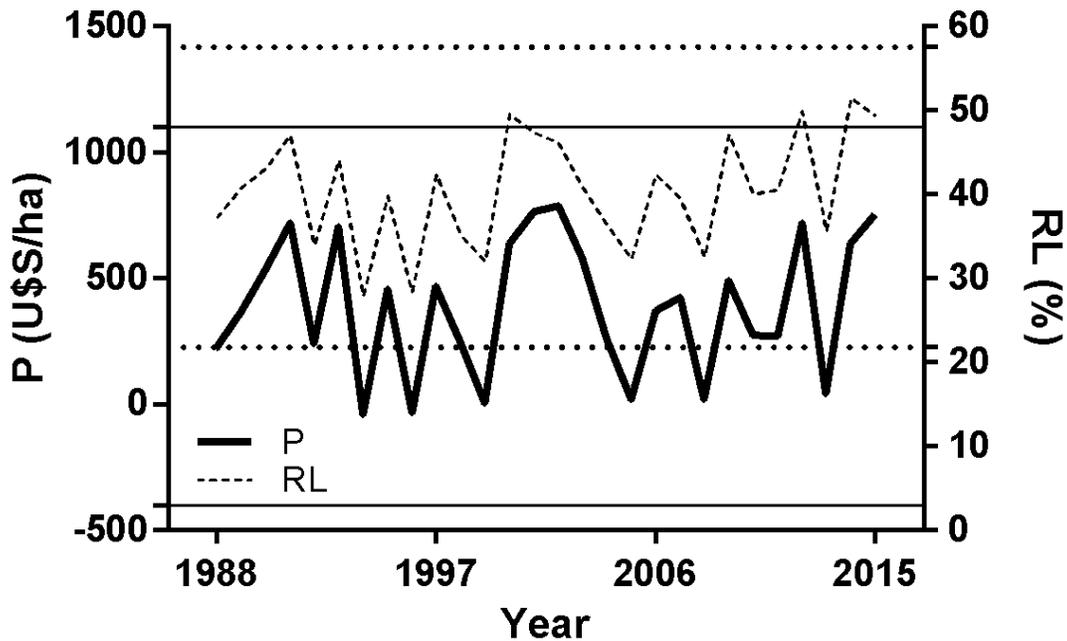

Figure 5. Landscape profit (P) and renewability (RL) for the Pergamino 1988-2014 simulation. The figure shows the average profit (P) and renewability level (RL) for each year during the Pergamino 1988-2014 simulation. Both P and RL have theoretical limits defined by the best or the worst conditions for both economic return (Profit) or environmental conditions (renewability of energy use). These limits are shown in solid and dotted lines for P and RL, respectively.

The interquartile P range (25–75% percentile) varied between 94 and 556 U$S / ha (Figure 6). However, the RL interquartile range was not as high as the observed in P (Figure 6). The relative homogeneity between each cropping production systems, in terms of external inputs, generated a remarkably low variability in RL terms during the whole simulated period. In this case, the interquartile RL range varied from 37.9% to 43.5 % with a median value of 41.2%, almost ten percentage points lower than the environmental threshold (ET) of 50% (Figure 6). While in terms of RL, the variability between agents was significantly lower than those observed for P, it was possible to detect both maximum and minimum RL values showing the model capability to simulate a wide area of decisions during the simulated period. Minimum average value for an agent outcome in

terms of RL was 29% and the maximum RL average value for an agent was 49.4% (Figure 6), a condition of very close agreement with the environmental threshold (ET). Agent behavior in AgroDEVS was also assessed by inspecting the intra-agent variability, using both P and RL during the simulation period (Figure 7). The interannual coefficient of variation of P showed a median value of 57%, with an interquartile range between 49.8% and 264.5 % (Figure 7). The RL interannual variation was significantly lower than P, exhibiting a median coefficient of variation of 18.9% and an interquartile range between 16 and 22% (Figure 7).

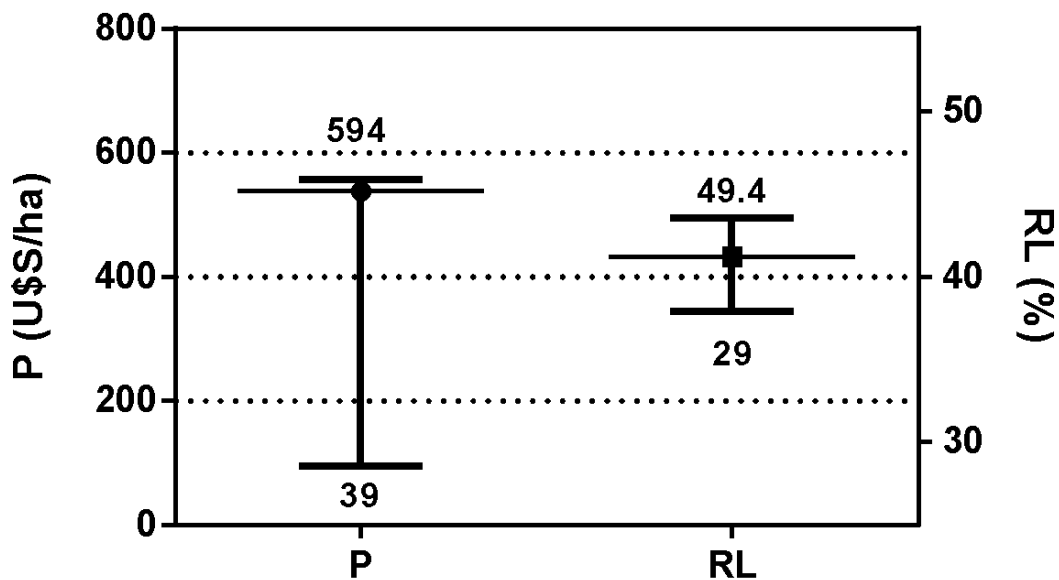

Figure 6 Inter-agent variability for the Pergamino 1988-2014 simulation. The figure shows the average profit (P) and renewability level (RL) of the agents in the Pergamino 1988-2014 simulation. The horizontal solid line shows the overall average of P and RL (n=625). The extremes of the whiskers represent the 25% and 75% quartiles, and the numbers show the minimum and maximum P and RL average values for all agents.

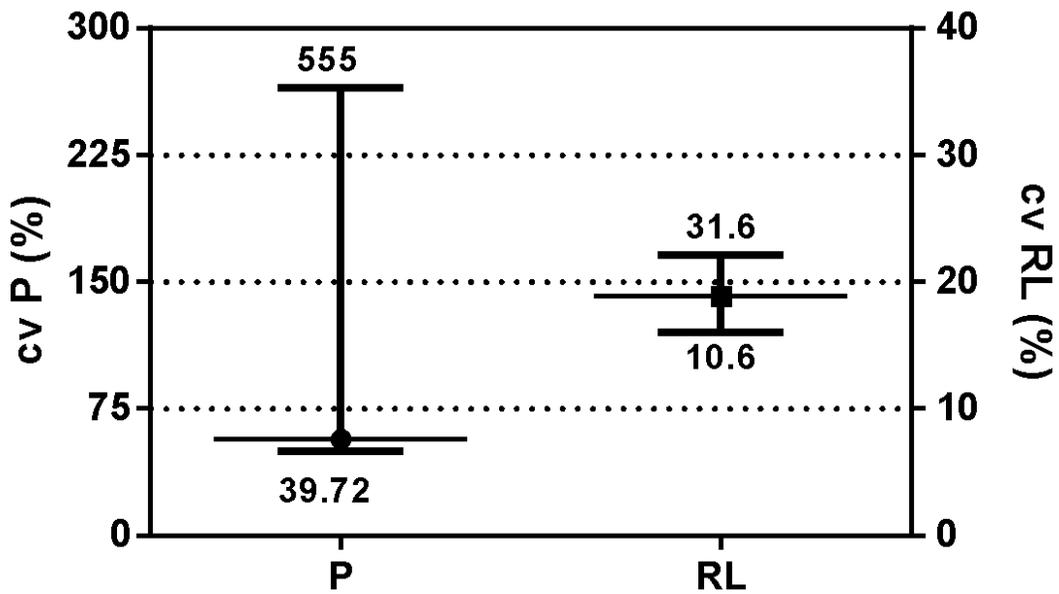

Figure 7. Intra-agent variability of the Pergamino 1988-2014 simulation. The figure shows the coefficient of variation of Profit (cv P) and Renewability Level (cv RL) for all 625 agents. The horizontal solid line shows the overall average of cv P and cv RL (n=625), the extremes of the whiskers represent the 25% and 75% quartiles, and the numbers denote the minimum and maximum cv P and cv RL values.

      A final assessment of the simulation results was carried for both the environmental (ET) and economic (AL) goal agreements exhibited by agents throughout the simulation period (Figure 8). The goal agreement metric indicates the percentage of years during which an agent fulfills each of the goals (environmental and economic). The agents exhibited an interquartile range of economic (AL) goal agreement between 32% and 64% with a median of 60% and maximum and minimum of 71.4% and 17.2%, respectively (Figure 8). These values are noticeably higher than the RL goal agreement that showed a median of 17.9% with an interquartile range between 7% and 25% (Figure 8). The agent's capability for adjusting the AL, based on both the current climate condition (WGC) and the AL fulfillment in the previous agricultural cycle, could explain the better goal agreement when compared against RL, which is a fixed parameter.

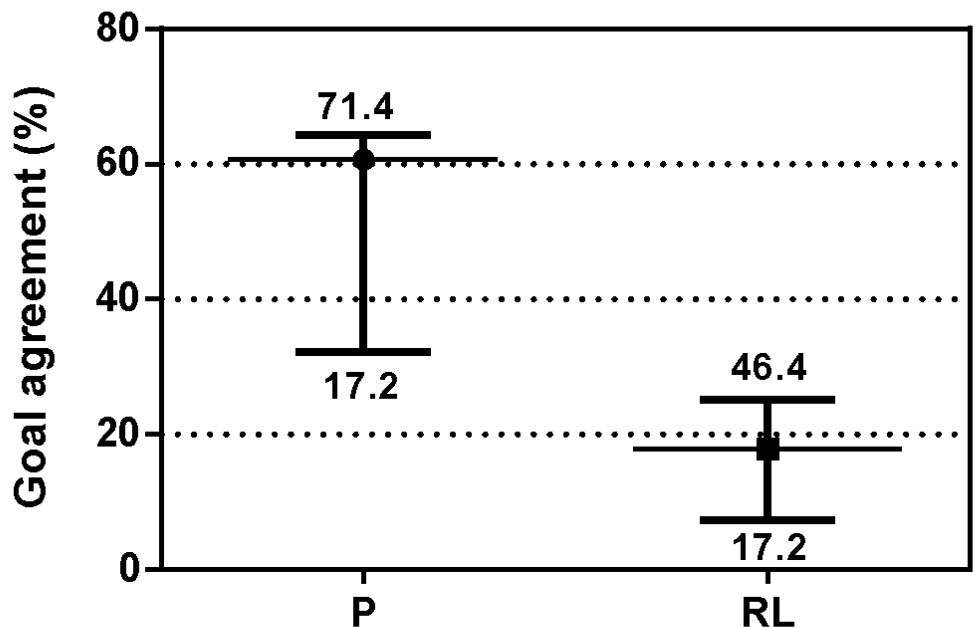

Figure 8. Ecological threshold (ET) and economic aspiration level (AL) goal agreements for the Pergamino 1988-2014 simulation. The figure shows the average percentage of agreement (% goal agreement) for all 625 agents of the Pergamino 1988-2014 simulation. The horizontal solid line shows the overall average of ET and AL goal agreement (n=625), the extremes of the whiskers represent the 25% and 75% quartiles, and the numbers show the minimum and maximum ET and AL goal agreement values.

When inspecting the extreme values for goal agreements, it was possible to identify a single agent exhibiting a maximum value of 46.4%, which means that, under the same economic and climate conditions, this agent was able to fulfill nearly 1 out of 2 years, the RL threshold (ET) by means of its crop type allocation decisions. Finally, the model outcome was also assessed by inspecting the final distribution of the three different technological levels (TL) across all 625 agents of the Pergamino simulation (Figure 9). The closely related observed and simulated agent distribution underline the model's ability to represent the process of agricultural intensification evidenced by a higher proportion of farmers using a technological level with higher levels of inputs. Both observed and simulated TL distribution patterns exhibit a dominance of the high TL. However, the model retains a greater percentage of agents at the lowest level (TL low) than in the observed data. It is possible that the model structure, which allows agents to remain using the lowest TL despite not reaching the minimum wealth to face those costs, is an explaining factor for the overestimation of Low TL at the end of the simulation.

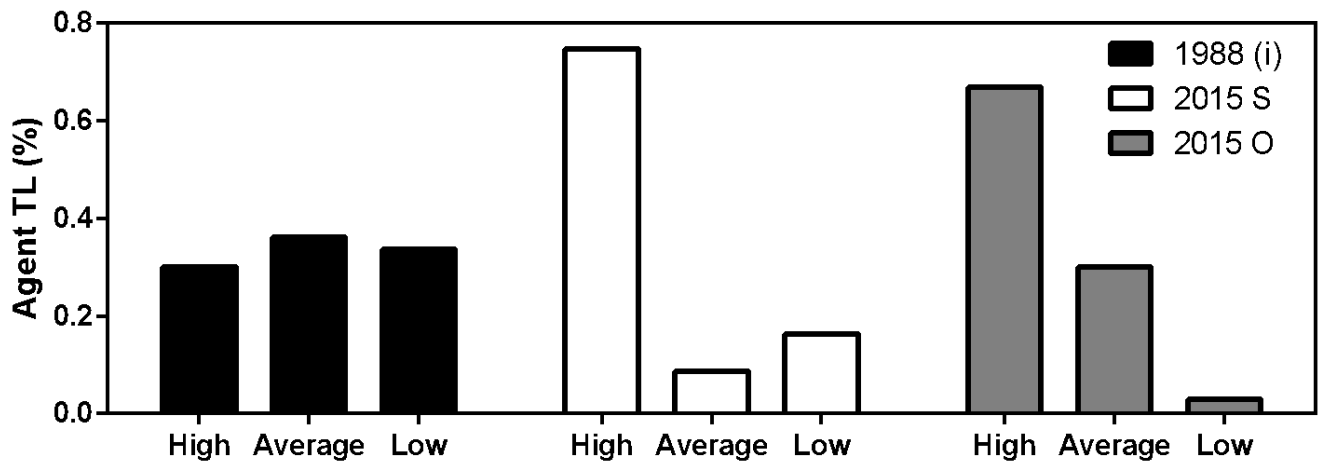

Figure 9. Initial (i) 1988 and final simulated (S) and observed (O) frequency values of agent technological level (Agent TL) among agents.

**Long-term scenarios**

*LUCC patterns*

The objective of the long-term simulations was to assess the magnitude of the effect of the tenure condition (i.e. an owner or tenant-dominated landscape) and climate (i.e. five contrasting climate regimes) on both LUCC (Figure 10), and P and RL outcomes (Figure 11). AgroDEVS simulations showed that the variability in climate regimes altered the pattern of crop type dominance at the end of the simulation cycles (Figure 10). Under constant climate regimes (Figure 10: L, A and H panels) the simulated landscape is always stabilized, in terms of LUCC, at higher values of soybean (S) cover, followed by the wheat/soybean double-cropping (W/S) and maize (M). On the other hand, under variable climate scenarios (Figure 10: V and R panels) the long-term simulations showed the highest crop type dominance represented by W/S instead of S, although M remained at the lowest cover throughout the simulated period. Tenure effect on LUCC dynamics was clearer under constant average climate regime (Figure 10: Panel A; 10O/90R). Under this scenario, the landscape dominated by tenants exhibited a much stronger S dominance, compared to W/S or M. Instead, under the same climate scenario, but dominated by owners (Figure 10: Panel A; 90O/90R), the model simulated a different LUCC dynamic, showing an earlier stabilization point for LUCC (ca. year 2) and very similar final cover values for the three analyzed crop types. Oppositely, the model showed similar LUCC dynamics in owners or tenants-dominated landscapes, under a constant unfavorable (Figure 10: Panel L) and constant favorable (Figure 10, Panel H) climate regimes. In variable climate scenarios, the regular see-saw climate change pattern (Figure 10, Panel V) showed less differences in LUCC pattern due to land tenure regimes with respect to the random climate pattern (Figure 10, Panel R). The inclusion in the simulation scenario of climate dynamics without a definite pattern (i.e. random) resulted in the increase of double cropping W/S dominance, and this effect was much stronger for the condition of a landscape dominated by owners, achieving in this condition the highest crop type dominance among the 10 long-term scenarios (Figure 10: Panel R; 90O/10R).

*Profit and Renewability level*

The variability between long-term scenarios was lower for both P and RL than for the simulated LUCC (Figure 11). Regarding the variability induced by the climate regime, the simulated landscapes were stabilized at increasing P-values, as climate scenarios were better (Figure 11: Panels L, A, and H). This effect of profit improvement occurred in both tenant- and owner-dominated landscapes. In this latter case, the stabilized P values were due to the differential income associated with non-payment of land rental.

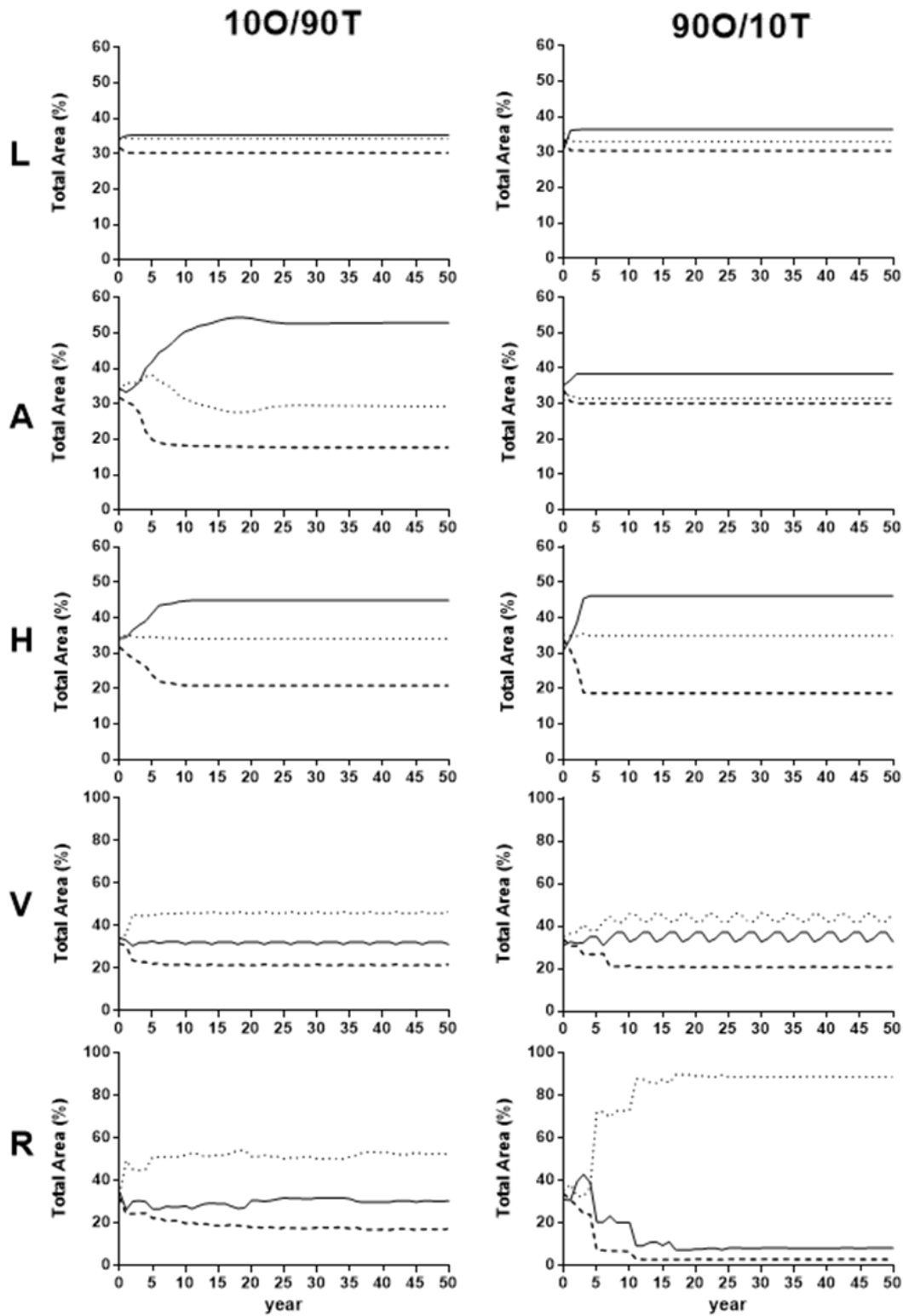

Figure 10. Simulated land use cover, expressed as % of Total Area, for the long-term simulations. The land uses are soybean (full line); double cropping wheat/soybean (dotted line) and maize (dashed line). The panels show the ten scenarios composed by land tenure regime (10O/90T: 10 % of owner agents and 90 % of tenant agents; and 90O/10T: 90 % of owner agents and 10% of tenant agents) and climate regime (L: constant unfavorable; A: constant average; H: constant favorable, V:

a see–saw pattern of very unfavorable-average-very favorable; and R: a random regime.

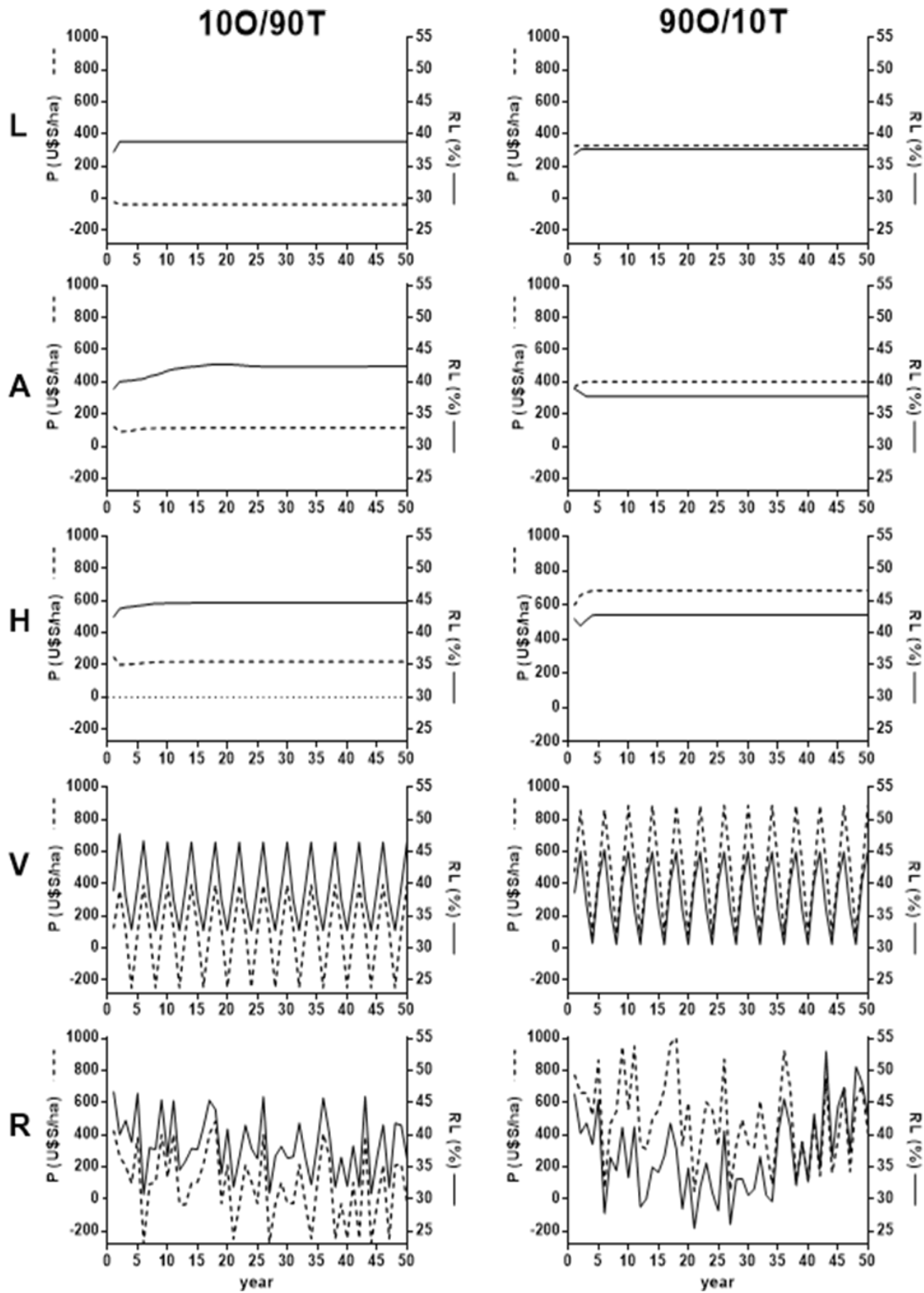

Figure 11. Simulated profit P (U$S/ha) and renewability level RL (%) for the long-term simulations. The panels show the ten scenarios composed by land tenure regime (10O/90T: 10% of owner agents and 90% of tenant agents; and 90O/10T: 90% of owner agents and 10% of tenant agents) and

climate regime (L: constant unfavorable; A: constant average; H: constant favorable, V: a see–saw pattern of very unfavorable-average-very favorable; and R: a random regime)

Although the simulated landscape configurations were clearly different (Figure 11), the RL variability among long-term scenarios under constant climate regimes (when stabilized) showed very small changes (less than 10%) between the maximum and minimum RL final values (Figure 11, panels L, A, V). When analyzing the scenarios under variable climate regime (Figure 11, panels V; R) a regular pattern of both RL and P variations was observed when climate evolve in a regular way (Figure 11, panel V). Moreover, the random climate regime (Figure 11, panel R) showed the highest interannual variability for both RL and P. Unlike what is observed in simulations under constant climate, in the case of variable climate scenarios, the model showed its sensitivity to weather growing conditions (WGC) in both RL and P, even after the LUCC stabilization. A clear evidence of this sensitivity is the highly variable RL and P simulated patterns under the random climate scenarios (Figure 11, panel R) even when the simulated landscape was stabilized in a configuration dominated by the wheat/soybean double-cropping (Figure 10, panel R).

## Discussion

Determining patterns of change in land use is undoubtedly a multifaceted challenge. When relying on simulations, several factors come into play such as climate influence on crop yield, the farmers' decisions, economic prices and costs, and the cognitive description of farmer behaviors (Hare and Deadman, 2004). Thus, the construction of a LUCC simulation model entails necessarily the coupling of social and environmental models (Müller et al., 2013). In this paper, we developed an ABM expressing several of these characteristics. AgroDEVS simulations were able to reproduce observable LUCC trends of the most representative cropping systems in the region under study. An ABM validation process has the peculiarity of being subjected to conflicts between achieving accuracy in matching the outcome of a simulation or in the processes simulated (Brown et al., 2005). This trade-off is usually solved based on the research goals. In the case of LUCC simulation models, both aspects are important. Predicting the LUCC trends is extremely important in decision making by policy makers (Verburg et al., 2004b). However, this information should be supplemented through an understanding of the underlying LUCC processes involved. This is required to identify potentially unsustainable land use regimes and correct them. AgroDEVS' results show that its structure is able to detect the overall trend on land use changes. This is done through clear and explicit modelling that reflects key process dynamics such as the climate influence on crop yields, farmer decisions and the landscape emergent properties due to farmer interactions at smaller scales. Moreover, there is always the chance to include other variables in the modelling exercise. In the interest of better representing the heterogeneity amongst different farmer's decision logics in AgroDEVS, future efforts could be aimed at exploring the farmers' decision-making process. This exercise could expose new relevant variables that improve the representation of agents' behavior. However, this inclusion would require a new setting and a more complex numerical validation. The cost-benefit balance of these additions should be analyzed carefully to avoid incurring in an overfitting, conspiring against the understanding of the true underlying phenomena under study (Brown et al., 2016).

   The contribution of the model developed in this work can be assessed by analyzing carefully the LUCC simulation results for Pergamino 1988-2015, while considering the trade-offs between output accuracy and processes understanding. Although the model was able to simulate the land use change dynamics of the three crops analyzed (i.e. the ordinal fit is always higher than 0.7), the adjustment based on the distance between observed and predicted (i.e. the v value) could be improved by the inclusion of other variables or exploratory processes (e.g. agricultural policy decisions not considered; the dynamics of prices and costs, etc.). However, the current model structure maintains the relative profit between activities which is highly sensitive to environmental conditions (WGC). Thus, the distance between the observed and predicted LUCC can be used as a predictor of the difference between what could have been a LUCC trend (based solely on the response to the environment in order to maximize profit, i.e. LUCC simulated) and another path that did not strictly follow the parameters of maximizing profit (i.e. LUCC observed). Notably, the region studied has been frequently subjected to decisions in agricultural policy (e.g. difficult marketing of some crops, imposition of export taxes) that influence strongly the changes in land use through mechanisms not directly related to profit activity (Porto and Lodola, 2013).

From the formal modelling and simulation point of view, the DEVS approach (and its related Cell-DEVS spatially specific flavor) offered several salient features, already discussed in model description section. Notably, in the context of socio-environmental systems, there is a feature that stands out. Namely, the input-output port-based hierarchical modularity permits the design of interdisciplinary models by composing complex systems through the interconnections of simpler ones. For instance, in AgroDEVS, the DEVS Atomic Model representing climate dynamics can be replaced by other more accurate or sophisticated DEVS model developed by a group of specialists in the climate domain, without requiring to alter the landscape (Cell-DEVS) portion of the system. This approach fits in the context of a current trend towards Systems of Systems-oriented modelling and simulation (Zeigler et al., 2012). This is particularly relevant for socio-natural-economic systems, whose domain-specific submodels are constantly subjected to revisions, improvements or replacements.

Incorporating environmental assessment in LUCC simulation models is a very desirable feature that is beginning to be explored (Veldkamp and Verburg, 2004). The analysis of environmental impacts on managed ecosystems has often been applied based reductionist approaches, identifying changes very accurately, but reducing the relevance of the results by not addressing an integrated or holistic approach for ecosystems modelling (Shanmuganathan et al., 2006). The AgroDEVS structure acknowledges the need for a systemic modelling by including emergy renewability level as a proxy for assessing ecosystem functioning of the cropping systems studied, something that has already been tested with agricultural systems both in the studied region and in other agricultural ecosystems (Agostinho et al., 2010; Dewulf et al., 2005; Ferraro and Benzi, 2013). Concerning the use of environmental work, the simulated ecosystems did not show a clear dynamic of increase or decrease in its reliance on energy from the economic system. Moreover, the results from the simulations, both in the Pergamino 1988-2015 and the long-term runs, also failed to show a clear trade-off between the environmental performance (assessed through renewability level) and economic performance (assessed through economic profit) in the studied systems. The simulated dynamics showed that the systems reduced their dependency on exogenous emergy (i.e. fossil energy) to the extent that environmental condition improved, and thus natural resources became more important. Clearly, this happens because at increasing levels of technology, increased use of external emergy is roughly proportional to the increase in capturing endogenous emergy associated with growth conditions (WGC) that improve crop yields. The high emergy return seems a characteristic of the farming systems of the Pampas region at the *field scale* (Ferraro and Benzi, 2015; Ferraro and Benzi, 2013); but the AgroDEVS simulations showed that this pattern is maintained at the *landscape scale* due to the emergent properties that arise from the integration of individual farmers behavior. However, in the context of changes in agricultural policy in the area under study, the results of AgroDEVS seem to indicate the need for policy options that alter the relative prices of crops, to avoid the predominance of monocultures or systems highly dependent on the soybean crop. In this context, the AgroDEVS capability for identifying potential tradeoffs between different agroecosystem domains (i.e. economic, environmental) is extremely important in the diagnosis of agricultural sustainability (Tittonell, 2014). Policy options, which have direct effects on model variables such as prices, can be readily tested into AgroDEVS for their repercussions on farm incomes and LUCC emergent patterns.

**Conclusions**

The model presented in this work represents an effort to integrate, within a complex system simulation, the effects of weather on crops, the farmer decision logic, and the cropping system profit as main LUCC drivers. AgroDEVS simulations based on real landscape data showed that LUCC direction was best represented than its magnitude, in terms of the land cover dynamics. The long-term simulations showed a dominance of cropping systems that included soybeans, and this dominance was stronger for monospecific soybean crops in scenarios under constant climate. The double cropping W/S that dominated mainly in scenarios under a variable climate. When assessing the farmer condition effect (i.e. tenure and climate) on LUCC, tenure resulted in much less effect on LUCC than the weather conditions (WGC). Finally, simulations showed no trade-offs between environmental and economic outcome both in simulation used to validate the model and long-term scenarios. The results suggest that LUCC modelling and its environmental and economic consequences is feasible and useful using an ABM approach. The AgroDEVS simulations allow not only to speculate on the LUCC dynamics, but to gain a greater understanding of the underlying processes involved. Future research should focus on improving the model structure to include different agent behaviors (e.g. multiple agent's profiles) as well as social and political factors both for predicting the LUCC direction, and to assess their relative magnitudes more accurately. Moreover, incorporating different productive regions of Argentina into AgroDEVS could expand the conclusions achieved in this paper, revealing differences in LUCC dynamics at an ecorregional level.

## Acknowledgements

This material is based upon work supported by the University of Buenos Aires (UBACYT-UBA); the National Council for Scientific Research (PIP-CONICET); and the National Agency for Science Promotion (PICT-ANPCyT) of Argentina. DB was supported by an undergraduate fellowship from FONSOFT-ANPCYT (JPT C2). SP was supported by a doctoral fellowship from CONICET.

# Appendix A. Sensitivity analysis

In a long-term simulation (28 years), we tested the effect of 1) crop prices, 2) WGC, 3) the Owner/Tenant agent ratio and 4) rental price. We run AgroDEVS varying one parameter at a time over a range of values (Table A1), while keeping the other parameters at their reference value (i.e. the initialization conditions of the Pergamino 1988-2015 simulation; for more details see Appendix B). The range of values for the crop prices was extracted from historical data and are the lowest and highest price of each crop for the years 2001-2020.

**Sensitivity analysis**

The WGC scenarios were created by alternating 50% of the campaigns with historical values (i.e. actual weather for that year) and the other 50% with the WGC level desired. Thus, five scenarios were created comprising 50% of historical values and 50% WGC (VU, U, R, F and VF). The owner/tenant agent relation ranged from 10% of the agents being owners and 90% tenants to 90% owners and 10% tenants in increments of 20% (i.e. 10/90, 30/70, 50/50, 70/30, 90/10). The rental price ranged from U$S/ha 221,6 to U$S/ha 775,6 (Table A1).

**Table A1.** Model parameters, units and range for sensitivity analysis. The reference value is related to the Pergamino 1988-2015 simulation

| Parameter | Units | Reference value | Range for SA |
|---|---|---|---|
| **Soybean Price** | U$S/tn | 277 | 141 - 346,4 |
| **Maize Price** | U$S/tn | 141 | 69,76 - 185,28 |
| **Wheat Price** | U$S/tn | 153 | 100.28 - 249,23 |
| **WGC** | - | - | 50% VU - 50% VF |
| **O/T agent ratio** | - | 63/37 | 10/90 - 90/10 |
| **Rental price** | U$S/ha | 443,2 | 221,6 - 775,6 |

The sensitivity analysis showed that the model is sensitive to wheat and soybean price changes in terms of profit. When the price of one of these crops rises, the land use cover that dominates the grid is W/S. This crop sequence has a high profit potential, and thus explains the change in profit observed. However, if the soybean price drops below the reference value the farmers cannot compensate with another LU and the profit drops. In terms of renewability, the model was not sensitive to changes in crop prices and all crops behaved similarly. Only a large increase in wheat price caused a 20% reduction in the RL (Figure A1).

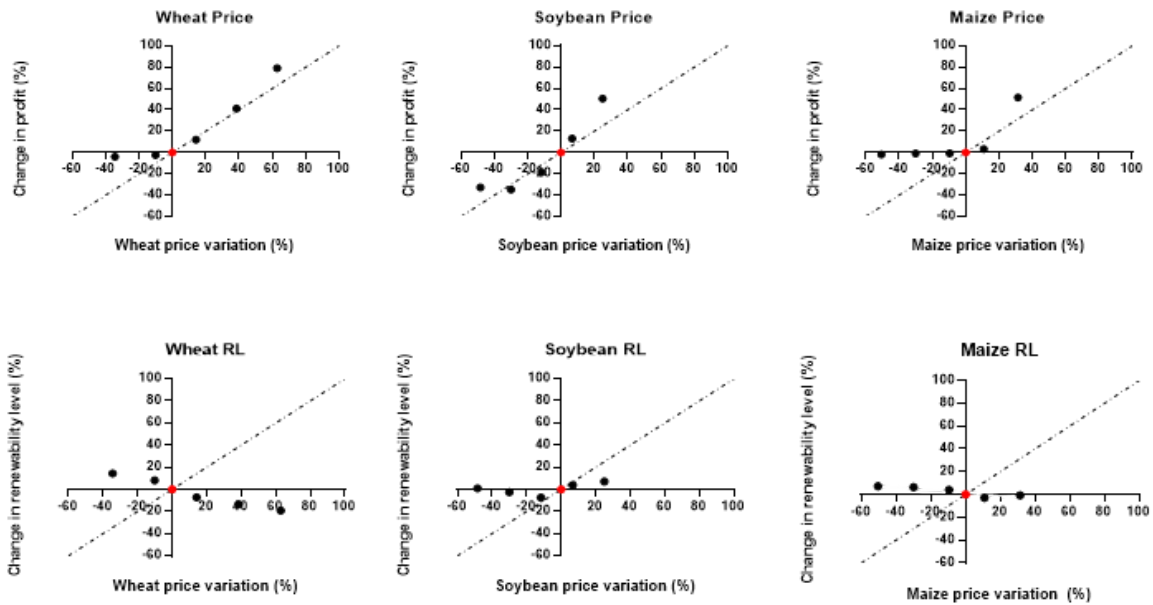

**Figure A1**. Changes in profit (above) and renewability (below) related to the Pergamino 1988-2015 simulation under increasing crop prices. Each dot shows the average value of the profit or renewability level of each simulation run with a different crop price. Profit's reference value (red dot) = 390.26 U$S/ha. Renewability level's reference value = 40.47 %. The dashed line is the line of identity.

The model's response to different weather patterns was clear in terms of change in profit (figure A2, left). Under favorable WGC, the profit was higher than the reference value. In the best-case scenario (50% of the years under VF weather), the profit was increased by 55% of the reference value. Conversely, under unfavorable WGC the profit tended to drop. In the worst-case scenario, (50% of the years under VU weather) the profit decreased 38%. This behavior is explained by the quick dominance of a more profitable LU (W/S) and management level (High) under favorable weather conditions, and the dominance of a defensive LU such as soybean under unfavorable weather conditions. The model showed little response in terms of RL under different

WGC (figure A2, right). The results indicate that under favorable weather conditions the RL could drop slightly due to the dominance of W/S. On the other hand, unfavorable weather conditions would lead to an increase in S, provoking an increase in RL.

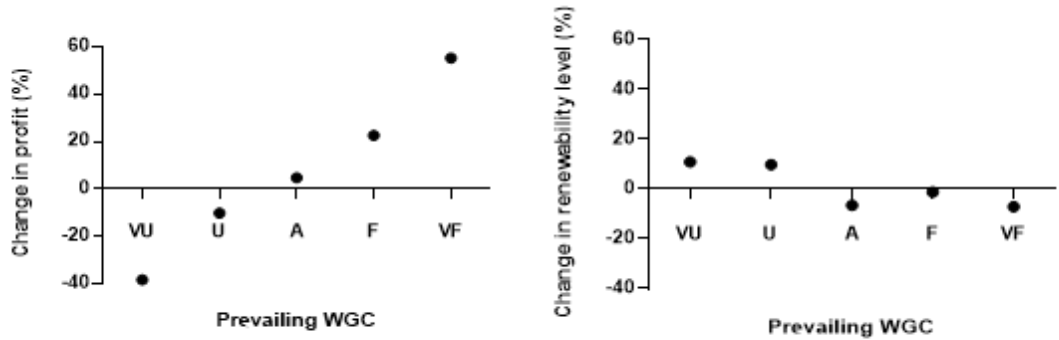

**Figure A2**. Changes in profit (left) and renewability (right) related to the Pergamino 1988-2015 simulation under different WGC scenarios. Each dot shows the average value of profit or renewability level of each simulation run with a different WGC. Profit's reference value = 390.26 U$S/ha. Renewability level's reference value = 40.47 %.

The model was slightly sensitive to changes in rental cost, leading to profit variation (figure A3, left). As expected, an increase in rental cost diminished the profit in tenants resulting in a decrease in the average landscape profit (no dominance of LU was observed here). On the other hand, the RL was not sensitive to changes in rental cost (figure A3, right). This could be explained by the relatively low fraction of the profit that the rental cost represents.

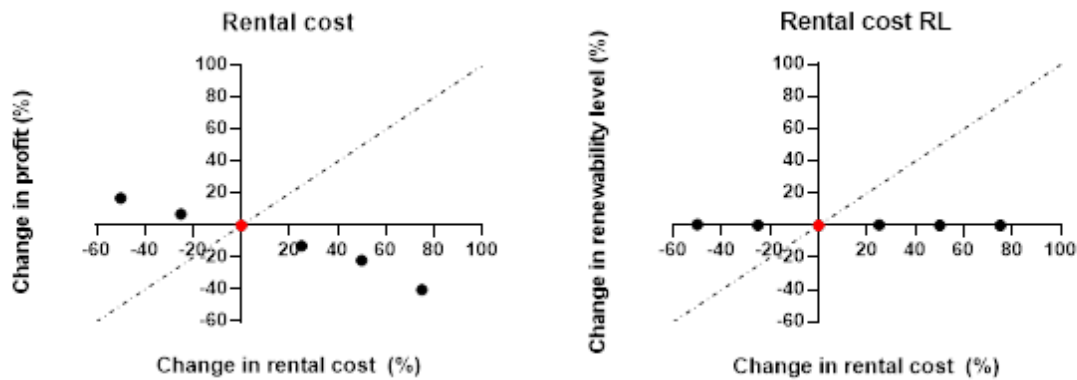

**Figure A3**. Changes in profit (above) and renewability (below) related to the Pergamino 1988-2015 simulation under increasing rental cost. Each dot shows the average value of profit or renewability level of each simulation run with a different rental cost. Profit's reference value (red dot) = 390.26 U$S/ha. Renewability level's reference value = 40.47 %. The dashed line is the line of identity

Regarding the O/T relation, the model was sensitive to changes in this parameter in terms of profit (figure A4, left). In the scenarios with a low O/T relation the average profit dropped, while with a high O/T relation the results showed an increase in profit. This behavior is explained by the increase in the number of farmers who pays rent and not by any LU or management changes. In terms of RL, the model was not sensitive to changes in the O/T relation (figure A4, right), supporting the results of the rental cost sensitivity analysis.

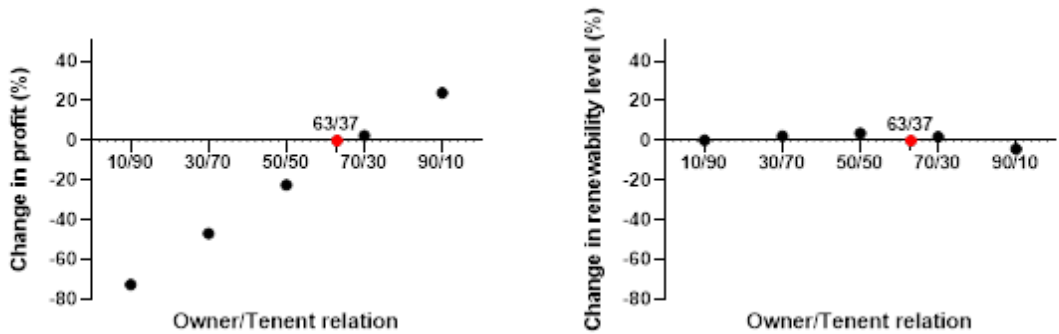

**Figure A4**. Changes in profit (left) and renewability (right) related to the Pergamino 1988-2015 simulation under increasing Owner/Tenant relation. Each dot shows the average value of profit or renewability level of each simulation run with a different Owner/Tenant relation. Profit's reference value (red dot) = 390.26 U$S/ha. Renewability level's reference value = 40.47 %.

**Appendix B. AgroDEVS initialization conditions.**

Crop yields are based on simulations using crop models in the Decision Support System for Agrotechnology Transfer (DSSAT) package (Jones et al., 2003) that have been calibrated for the studied location (Mercau et al., 2007). Crop management of each TL (e.g. genotype, sowing date, fertilizer rate, pesticide use, tillage operations, and production costs) was used for running the DDSAT crop yield simulations, and was based on local management data. Crop yield variability during the DSSAT simulated period (1971-2008) was used to obtain the crop yield under contrasting WGC levels. The five WGC levels (very unfavorable, unfavorable, average, favorable, and very favorable) correspond to different percentiles (i.e. 10, 30, 50, 70, and 90, respectively) of yields simulated using historical weather records. DSSAT simulations do not account for important factors such as weeds, diseases, and pests. Thus, we empirically adjusted the attainable crop yield (van Ittersum and Rabbinge, 1997) resulting from DSSAT simulations in order to model actual crop yield. Local data of simulated versus observed crop yield were used for obtaining the adjusting coefficients (attainable to actual yield) at each TL for each crop species (Mercau et al., 2001; Mercau et al., 2007; Satorre et al., 2005). AgroDEVS reflects three incremental TL: low (L), average (A) and high (H) technological levels; and three crop types (maize, soybean and wheat/soybean double-cropping, see region under study and cropping system description section for production system description). Using DSSAT and historical records of production costs, two of the look-up tables are built: 1) a crop yield matrix (Table B1) and 2) a cost matrix for representing the full crop type (3 levels), TL (3 levels) and WGC (5 levels) combination (Table B2). As for crop yields and production costs, the emergy accounting method is used for building a third look-up table for representing the full combination of crop types (3 levels), TL (3 levels) and WGC (5 levels) (Table B3). Historical prices of maize, soybeans, and wheat, as well as production costs (i.e. fertilizers, seeds, pesticides, and harvest and sale costs), were extracted from the Argentine trade magazine "Márgenes Agropecuarios" (http://www.margenes.com). In all scenarios (i.e. the

Pergamino simulation and the long-term simulations) we assumed constant output prices equal to median for 2008–2015.

**Table B1:** Simulated crop yields using DSSAT (expressed in t/ha) for the combinations of WGC, LU and TL levels. Land use (LU) classes: Maize; Soybean; Wheat/Soybean double cropping. Technological level (TL) classes: H (High); (A) Average; L (Low); Weather Growing Condition (WGC) classes: Very Unfavorable; Unfavorable; Average; Favorable; Very Favorable.

| LU | TL | WGC | | | | |
|---|---|---|---|---|---|---|
| | | Very Unfavorable | Unfavorable | Average | Favorable | Very Favorable |
| Maize | L | 4,05 | 6,27 | 7,45 | 8,37 | 9,25 |
| | A | 4,88 | 7,78 | 9,02 | 10,45 | 11,59 |
| | H | 5,40 | 8,80 | 10,22 | 11,94 | 13,18 |
| Soybean | L | 1,89 | 2,67 | 3,13 | 3,72 | 4,15 |
| | A | 2,13 | 3,00 | 3,53 | 4,18 | 4,67 |
| | H | 2,37 | 3,34 | 3,92 | 4,65 | 5,19 |
| Wheat/ Soybean | L | 3,06 | 4,34 | 5,21 | 5,73 | 7,11 |
| | A | 3,53 | 4,89 | 6,01 | 6,55 | 8,16 |
| | H | 4,25 | 5,85 | 7,30 | 7,90 | 9,83 |

**Table B2:** Crop production costs (expressed in U$S/ha) for the combinations of WGC, LU and TL levels. Land use (LU) classes: Maize; Soybean; Wheat/Soybean double cropping. Technological level (TL) classes: H (High); (A) Average; L (Low); Weather Growing Condition (WGC) classes: Very Unfavorable; Unfavorable; Average; Favorable; Very Favorable.

| LU | TL | WGC | | | | |
|---|---|---|---|---|---|---|
| | | Very Unfavorable | Unfavorable | Average | Favorable | Very Favorable |
| Maize | L | 504 | 619 | 680 | 727 | 773 |
| | A | 618 | 768 | 832 | 906 | 965 |
| | H | 717 | 892 | 966 | 1055 | 1119 |
| Soybean | L | 262 | 302 | 326 | 356 | 378 |
| | A | 329 | 374 | 401 | 435 | 460 |
| | H | 395 | 446 | 476 | 514 | 541 |
| Wheat/ Soybean | L | 477 | 511 | 528 | 541 | 584 |
| | A | 618 | 656 | 675 | 690 | 738 |
| | H | 759 | 801 | 822 | 838 | 892 |

**Table B3:** Crop renewability values (expressed in %) for the combinations of WGC, LU and TL levels. Land use (LU) classes: Maize; Soybean; Wheat/Soybean double cropping. Technological level (TL) classes: H (High); (A) Average; L (Low); Weather Growing Condition (WGC) classes: Very Unfavorable; Unfavorable; Average; Favorable; Very Favorable.

| LU | TL | WGC t | | | | |
| --- | --- | --- | --- | --- | --- | --- |
| | | Very Unfavorable | Unfavorable | Average | Favorable | Very Favorable |
| Maize | L | 35,5 | 40,2 | 42,0 | 45,8 | 50,2 |
| | A | 33,1 | 37,8 | 39,6 | 43,4 | 47,8 |
| | H | 31,0 | 35,6 | 37,3 | 41,2 | 45,6 |
| Soybean | L | 43,7 | 48,4 | 50,1 | 53,6 | 57,6 |
| | A | 42,2 | 46,9 | 48,7 | 52,3 | 56,3 |
| | H | 40,8 | 45,5 | 47,3 | 50,9 | 55,1 |
| Wheat/ Soybean | L | 24,3 | 28,3 | 29,9 | 33,4 | 37,5 |
| | A | 23,0 | 26,9 | 28,5 | 31,9 | 36,0 |
| | H | 21,8 | 25,7 | 27,2 | 30,5 | 34,7 |

**Table B4:** Median output (crop) prices for the 2008–2015 period

| Land uses (LU) | Price (U$S/tn) |
| --- | --- |
| Maize | 141 |
| Soybean | 277 |
| Wheat/ Soybean | 153 |

**Table B5:** Adjustment factor αAL(WGC) of the aspiration level (AL) due to weather growing condition (WGC) level. WGC classes: Very Unfavorable; Unfavorable; Average; Favorable; Very Favorable.

| WGC | | | | |
| --- | --- | --- | --- | --- |
| Very Unfavorable | Unfavorable | Average | Favorable | Very Favorable |
| -0.55 | -0.28 | 0.00 | 0.22 | 0.45 |

**Table B6:** Adjustment factor αAL(BN) of the aspiration level (AL) due to agent technological level (TL) and the best neighbor technological level $TL_t$ (BN). A positive (or negative) αAL(BN) value indicates an increase (or decrease) in AL since the best neighbor exhibits a higher (or lower) TL than that evaluated by the agent. For equal TL values, αAL(BN) = 0. TL classes: H (High); (A) Average; L (Low)

| TL t (BN) | TL t | | |
| --- | --- | --- | --- |
| | L | A | H |
| L | 0.00 | 0.20 | 0.45 |
| A | -0.25 | 0.00 | 0.20 |
| H | -0.55 | -0.25 | 0.00 |

**Table B7:** Working capital threshold (WCT) for each technological level (TL) class (expressed in U$S/ha.). WCT values equal to 60% of the highest production cost for each TL, considering and average indebtedness rate of 40 % (AACREA 2014)

| TL | WCT (U$S/ha) |
|---|---|
| Low | 252 |
| Average | 333 |
| High | 413 |

## Appendix C. Submodels description

*Profit calculations*

P is calculated as the gross income (yield times product price) minus direct costs. Direct production costs include fixed and variable components. Fixed direct costs do not depend on crop yield (e.g., seed and agrochemicals). Oppositely, variable direct costs are yield-dependent (e.g., harvest, marketing fees and grain transportation). Finally, each farmer P is calculated using the individual crop profit affected by each crop type allocation into the agent farm. Figure C1 is an excerpt of the rule that calculates P: an agent has a defined LU and TL and is capable of sensing the WGC, and therefore it can calculate its P by multiplying its crop type allocation, yields and prices, and subtracting costs.

```
                    rule:
Values assigned ←  {~pro := (((0,0)~lu1/ 100) *    ((#macro(yield_lu1) *
                    #macro(price_lu1)) -  #macro(cost_lu1))) +

                    (((0,0)~lu2/ 100) * ((#macro(yield_lu2) * #macro(price_lu2)) -
                    #macro(cost_lu2))) +

                    (((0,0)~lu3/ 100) * (((#macro(yield_lu4) * #macro(price_lu4)) +
                    (#macro(yield_lu5) *

                    #macro(price_lu2))) - #macro(cost_lu3))) - ((0,0)~rent *
                    (#macro(price_lu2)))}
                    ...
Delay for assignments ← {0}
Conditions of this rule ← { (0,0)#macro(climateReceived)       and
                    (not isUndefined((0,0)~lu1))               and
                    (not isUndefined((0,0)~lu2))               and
                    (not isUndefined((0,0)~lu3))               and
                    (not isUndefined((0,0)~mgm)) }
```

Figure C1. Excerpt of the rule that performs agents' profit calculations. In the cases of yield, price and cost, #macro involves lookup tables.

*Renewability level calculations*

AgroDEVS calculates Renewability Level (RL) values by using the emergy synthesis procedure (Odum, 1996). Briefly, the emergy accounting methodology tries to account for both the natural and human-made capital storages. The emergy accounting method values these storages using a common unit of reference, the solar equivalent joule (seJ). The method accounts for the environmental support provided directly and indirectly by nature to resource generation and processing; it focuses on valuation of the intrinsic properties of ecosystems (Mellino et al., 2015). For further details on emergy synthesis methodology see (Brown et al., 2001) and (Ferraro and Benzi, 2015). AgroDEVS uses the renewability level as a sustainability metric (Giannetti et al., 2010). RL is then calculated as the ratio of renewable emergy to total emergy use, as it follows:

RL (%) = R / (R + N + F + S)

where,

RL (%): renewability level value

R (sej / ha$^{-1}$ year$^{-1}$): renewable flows from nature

N (sej / ha$^{-1}$ year$^{-1}$): nonrenewable flows from nature

F (sej / ha$^{-1}$ year$^{-1}$): imported economic flows

S (sej / ha$^{-1}$ year$^{-1}$): services

Therefore, at each model time step (i.e. a CC) each individual farmer renewability level (RL) is used as a measure of environmental impact. In the figure below (Figure C2), an agent uses its LU and TL along with the sensed WGC to obtain the final farmer's RL value (by affecting the individual crop renewability level with the crop type allocation into the agent's farm).

```
                         rule:
                         ...
    Values assigned ←─ {~eme :=  (((0,0)~lu1/ 100) * #macro(emergy_lu1)) +
                         (((0,0)~lu2/ 100) * #macro(emergy_lu2))          +
                         (((0,0)~lu3/ 100) * #macro(emergy_lu3));}
                         ...
  Delay for assignments ←─ {0}
  Conditions of this rule ←─ {(0,0)#macro(climateReceived)        and
                         (not isUndefined((0,0)~lu1))             and
                         (not isUndefined((0,0)~lu2))             and
                         (not isUndefined((0,0)~lu3))             and
                         (not isUndefined((0,0)~mgm)) }
```

Figure C2. Excerpt of the rule that performs agents' renewability calculations. In this case, #macro (emergy) involves emergy lookup tables.

*Agent decisions*

Both P and RL values for each agent are related to the AL and ET thresholds, in order to trigger the agent's decisions. In AgroDEVS, the economic goal (AL) is dynamic and it is based on the aspiration level adjustment. It represents the currently dominating economic paradigm, where the receiver (the agent) is the market actor who decides the system outcome value (Grönlund et al., 2015). Oppositely, the environmental threshold (ET) is fixed and it represents a strong sustainability view, where the value approach is grounded in systems science rather than economic science, where a value focused on the system level is accepted (Grönlund et al., 2015). During the model simulation process, the fulfillment of the economic goal drives the crop type allocation. Thus, the LUCC process is triggered when the economic threshold (i.e. aspiration level) is not accomplished.

<u>*Aspiration level adjustment*</u>

A first AL adjustment is based on the WGC, as it follows:

$$CAL_t = AL_t + AL_t * \alpha AL(WGC)$$

where

$CAL_t$ = climate-adjusted aspiration level (U$S/ha)

$AL_t$ = aspiration level (U$S/ha) calculated at the end of the previous CC

$\alpha AL(WGC)$ = adjustment factor of AL due to WGC level for the current CC (see Appendix B, Table B5 for $\alpha AL(WGC)$ values used in the simulations).

A second AL adjustment defines the aspiration level for the next CC and it is based on the learning and adaptation model (Bert et al., 2011). If $P_t > CAL_t$ then the next aspiration level is subjected to an incremental adjustment using a weighted average, as it follows:

$$AL_{t+1} = 0.45\ AL_t + 0.55\ P_t$$

where

$AL_{t+1}$ = aspiration level (U$S/ha) for the next CC

$CAL_t$ = aspiration level (U$S/ha) after current CC adjustment

$P_t$ = Profit (U$S/ha) calculated in profit calculations section

```
rule:
...
Values assigned ← {~AL_threas  :=  if ($aju = 0, (0,0)~AL_threas, 0)      +
                   if ($aju = 1,  if (((0.45 * ((0,0)~AL_threas          +
                   ((0,0)~AL_threas * #macro(adjust_climate)))) + (0.55 * (0,0)~pro))
                   > 0, (0.45 *     ((0,0)~AL_threas + ((0,0)~AL_threas *
                   #macro(adjust_climate)))) + (0.55 * (0,0)~pro), 0), 0)
                   ... }
Delay for assignments ← {0}
Conditions of this rule ← { (0,0)#macro(Processing) and #macro(neighborsProcessing) }
```

Figure C3. Excerpt of an agent's aspiration level adjustment. In this example, the agents' profit was larger than its AL, so it proceeds to perform an incremental adjustment.

In the figure above we can see an excerpt of the rule for adjusting incrementally the aspiration level using a weighted average when an agent's $P_t > CAL_t$. When both the agent and its neighbors finish with the calculation of their P, RL and the fulfillment of both economic and environmental goals, the agent proceeds to adjust its AL for the next cropping cycle. When the economic outcome is lower than the aspiration level (i.e. $P_t < CAL_t$), the agent's perception is extended to include the influence of the physical (Moore) neighbors, and the $AL_{t+1}$ is adopted by inspecting the neighbors' P outcomes. Thus, the farmer adopts $AL_{t+1}$ using the neighbors' P data. If there is at least one neighbor under the condition of $P_t > CAL_t$ then the farmer selects its best neighbor (BN), in profit terms, and the $AL_{t+1}$ is calculated using both the AL for the best neighbor, $CAL_t$ (BN), as well as an adjustment factor due to differences in TL between farmers:

$$AL_{t+1} = CAL_t (BN) + CAL_t (BN) * \alpha AL(BN) \qquad (4)$$

where:

$AL_{t+1}$ = aspiration level (U$S/ha) for the next CC

BN = the agent with the highest P value in the Moore neighborhood (n = 8)

$CAL_t$ (BN) = aspiration level (U$S/ha) for the BN in the current CC

αAL(BN) = adjustment factor of AL due to BN technological level (see Appendix B, Table B6 for αAL(WGC) values used in the simulations).

In the case that $P_{BN}$ <= $CAL_t$ (i.e. no neighbor meets the economic threshold) then the next aspiration level is subjected to a detrimental adjustment using a weighted average, as it follows:

$$AL_{t+1} = 0.55\ CAL_t + 0.45\ P_t \qquad (5)$$

where:

$AL_{t+1}$ = aspiration level (U$S/ha) for the next CC

$CAL_t$ = aspiration level (U$S/ha) after current CC adjustment

$P_t$ = Profit (U$S/ha) calculated in *Profit calculations*

Different values for incremental and detrimental adjustment (0.45 and 0.55 respectively) was applied in order to simulate the farmer willing to tolerate higher payoffs more rapidly than lower ones, thus showing greater resistance to downward changes (Gilboa and Schmeidler, 2001).

*Technological level adjustment*

Farmers are also able to upgrade or downgrade its technological level (TL) after inspecting its own P outcome in the CC. The rules for adjusting the TL are based on the working capital threshold (WCT) which represent the highest production cost within each TL (Table B7). Rules for TL adjustment are:

$$TL_{t+1} = H \text{ if } P_t > WCT\ (H) \qquad (6)$$

$$TL_{t+1} = A \text{ if } P_t > WCT\ (A) \text{ and } P_t < WCT\ (H) \qquad (7)$$

$$TL_{t+1} = L \text{ if } P_t < WCT\ (A) \qquad (8)$$

where,

$TL_{t+1}$ = Technological level for the next CC (H: high; A: average and L: low)

WCT (Tuli): working capital threshold (average or high) for TLi (low, average and high)

```
rule:
...
Values assigned ← {~mgm :=    if (#macro(MGM_adaptive) = 1,
                  if ((0,0)~pro > #macro(wc_max_mgm_3), 3,
                  if ((0,0)~pro > #macro(wc_max_mgm_2), 2,
                  1)     ), (0,0)~mgm);
                  ...}
Delay for assignments ← {0}
Conditions of this rule ← {(0,0)#macro(Processing) and #macro(neighborsProcessing) }
```

Figure C4. Excerpt of an agent's technological level adjustment. In this case, #macro (wc_max) involves WCT lookup table (Appendix B, Table B7).

In Figure C4 we can see an excerpt of the rule for adjusting an agent's TL. When both the agent and its set of neighbors are done with calculating their P, RL and evaluating the fulfillment of both economic and environmental goals, the agent proceeds to adjust its TL for the next cropping cycle by comparing its P with the WCT.

*Land use and cover change (LUCC)*

Final farmer decision defines the land use configuration. The LUCC process is triggered when the economic threshold is below the economic outcome and also there is at least one neighbor in the Moore neighborhood that meets the economic threshold as follows:

$P_t < CAL_t$ and $P_t (BN) > CAL_t$

where,

$P_t$ = Profit (U$S/ha) calculated as in profit calculations section

$P_t (BN)$: Best Neighbor Profit (U$S/ha)

$CAL_t$ = Aspiration level (U$S/ha) after current CC adjustment

After inspecting this condition, the agent selects the crop type allocation of the best neighbor (BN) as follows:

$$LU\ i_{t+1} = LU\ i_t\ (BN)$$

where,

$LU\ i_{t+1}$: Percentage of agent's farm area under crop type i for the next CC (i = corn, soybean or wheat/soybean)

BN = the agent with the highest P value in the Moore neighborhood (n = 8)

$LU\ i_t(BN)$: Percentage of BN farm area under crop type i for the next CC

```
                          rule:
                          ...
Values assigned ←─ {$clu       := 1;
                    $aju       := 3;
                    $clu1      := (1,0)~lu1;
                    $clu2      := (1,0)~lu2;
                    $clu3      := (1,0)~lu3;
                    $cAL_threas := (1,0)~AL_threas;
                    $cmgm      := (1,0)~mgm; }
Delay for assignments ←─ {0}
Conditions of this rule ←─ { (0,0)#macro(ParametersCalculated)       and
                    #macro(neighborsParametersCalculated)  and
                    (not isUndefined((1,0)~pro))           and

                    ((isUndefined((1,-1)~pro)  or  (1,-1)~pro   <= (1,0)~pro)   and
                    (isUndefined((0,-1)~pro)   or  (0,-1)~pro   <= (1,0)~pro)   and
                    (isUndefined((-1,-1)~pro)  or  (-1,-1)~pro  <= (1,0)~pro)   and
                    (isUndefined((-1,0)~pro)   or  (-1,0)~pro   <= (1,0)~pro)   and
                    (isUndefined((-1,1)~pro)   or  (-1,1)~pro   <= (1,0)~pro)   and
                    (isUndefined((0,1)~pro)    or  (0,1)~pro    <= (1,0)~pro)   and
                    (isUndefined((1,1)~pro)    or  (1,1)~pro    <= (1,0)~pro)
```

Figure C5. Excerpt of the rule that copies an agent's neighbor's land use, aspiration level and technological level. These attributes are not yet assigned to the agent.

In the figure above there is an excerpt of the rule that copies the best neighbor's land use allocation. If both an agent and its neighbors have already calculated their P and RL then the model will proceed to select the neighbor with the highest P and copy its LU allocation, AL and TL to be used later in the simulation and, if necessary, assigned to the agent.

# Appendix D. Input data

## Pergamino 1988-2015 simulation

Input data from 1988-2015 were used for both the model initialization and to test the model outcomes against the actual land use changes in the studied area. Table D1 presents a summary of these initial conditions.

Table D1. Summary of the initialization conditions for the Pergamino 1988-2014 simulation. Land use classes: M (Maize); S (Soybean); W/S (Wheat/Soybean). Technological level classes: H (High); (A) Average; L (Low). WCT is the Working capital threshold for each technological level class. Descriptor value represents the type of the model component: P (parameter: fixed); A (attribute: variable during simulation due to ABM rules). Condition marked with an asterisk represent the real data extracted from the 1988 National agricultural census (INDEC 1991). Values of cropping regime (cost, renewability, yield, prices) as well as AL adjustment factor and WCT are shown in the Appendix B. RP represents the value of 1.6 t of Soybean crop.

|  | Condition | | | | | | |
|---|---|---|---|---|---|---|---|
|  | Number of agents | Owner / Tenant ratio* | Crop type allocation at landscape level (%) * | Technological level agent distribution al landscape level (%) * | Aspiration level (U$S/ha) | Ecological threshold (%) | Rental price (U$S) |
| **Symbol** | # agents | O/T | LU | TL | AL | ET | RP |
| **Value** | 625 | 63/37 | 20 (M)<br>36.2 (S)<br>35.8 (W/S) | 30 (H)<br>36 (A)<br>32 (L) | 0.6 WCT | 50 | 443.2 |
| **Descriptor** | P | P | A | A | A | P | C |

The observed LUCC dynamics were obtained from agricultural surveys (SAGPyA, 2009), and both the initial (1988) and final (2015) TL frequency distribution among farmers were obtained from the national agricultural census (BOLCER, 2015; INDEC, 1991). In order to represent the main tenure regimes in the studied area (owners and tenants), AgroDEVS was initialized using the 1988 Owner/Tenant relation (INDEC, 1991). The land rental price was set to historical values of 1.6 t of soybean per hectare (Margenes_Agropecuarios, 2015). Crop yield and renewability level values under different weather conditions (WGC) and technological levels (TL) were calculated as it is explained in Appendix C profit calculations section. In the initial landscape configuration, each farmer was assigned an initial working capital (WC) according to his initial TL. Initial AL level were fixed for each farmer in order to account for the 60% of the direct costs (i.e. 0.6 WCT) of the

TL adopted in the first CC of the simulation (AACREA, 2014). Oppositely, the ET value remains constant during the simulations and is fixed in RL = 50% (Table 1). The accuracy of the simulations was assessed using both 1) the squared distances between a simulation's outputs and a set of observations (RMSE) and 2) the ordinal pattern analyses (OPA) (Thorngate and Edmonds, 2013). OPA indicates the topological fit between observed and simulated outputs but does not consider their closeness.

**Long-term scenarios**

AgroDEVS was also run over a 50-year period, under contrasting scenarios due both 1) five WGC regimes (constant unfavorable, constant average, constant favorable, a see–saw pattern of very unfavorable-average-very favorable, and a random regime), and 2) two tenure regimes based on the landscape pattern of Owner/Tenant agent relationship (90/10 and 10/90). As tenants are more focused on short-term income and are less likely to invest in longer-term management strategies than owners (Soule et al., 2000), we used AgroDEVS for testing the hypothesis that tenant farmers are less likely than owner-operators to adopt crop allocation decisions that lead to sustainable LUCC trajectories. All crop types and TL were set to equal distribution in the landscape (i. 33% of the total area for each crop and each TL) at the initialization. However, the internal assignment of each crop type for each agent was set randomly. Model simulations were inspected in terms of the dynamics of 1) crop type allocation of total area, 2) profit and 3) renewability level.

**Appendix E. Modelling approach**

In Figures E1 and E2 we provide an illustrative, non-exhaustive sample view of the Cell-DEVS specification used for AgroDEVS. Most of the information declaring the structure, components and behavior is found in a text file with .ma extension.

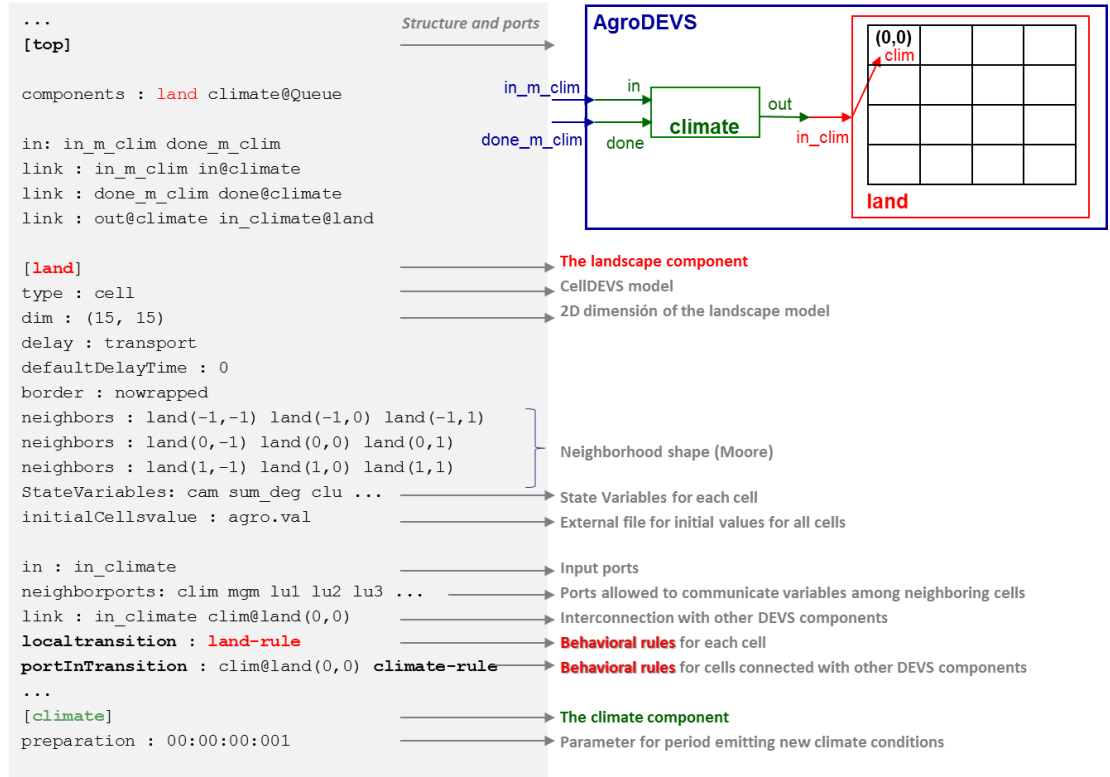

Figure E1. Sample excerpt of an AgroDEVS.ma definition file (Left) Schematic of the AgroDEVS system. The "climate" DEVS model is connected with the "landscape" Cell-DEVS model (Right).

In Figure E1 we can see the main statements defining structure, components and interconnections through input-output ports. In Figure E2 we provide an excerpt of the agrodevs.ma file highlighting relevant lines. Note that three dots "…" in the figures denote lines omitted for the sake of brevity. In Figure E2 we can see main statements defining behavior for the landscape cellular automaton. Within each cell, a list of rules is evaluated sequentially in a top-down fashion, using a *Value*, *Delay*, *Condition* structure: the first rule that evaluates its *Condition* to true, will apply the *Value* to its attributes after a *Delay* amount of time. Once a rule is applied, the simulator recommences the cycle evaluating the rules from the top. This mechanism is applied asynchronously, simultaneously and in parallel to all cells in the model. The global timing for the whole cellular space is an emergent property driven by the local timing applied by each independent cell. The pairs (X,Y) denote the positions for neighbors relative to each currently evaluated cell denoted with (0,0).

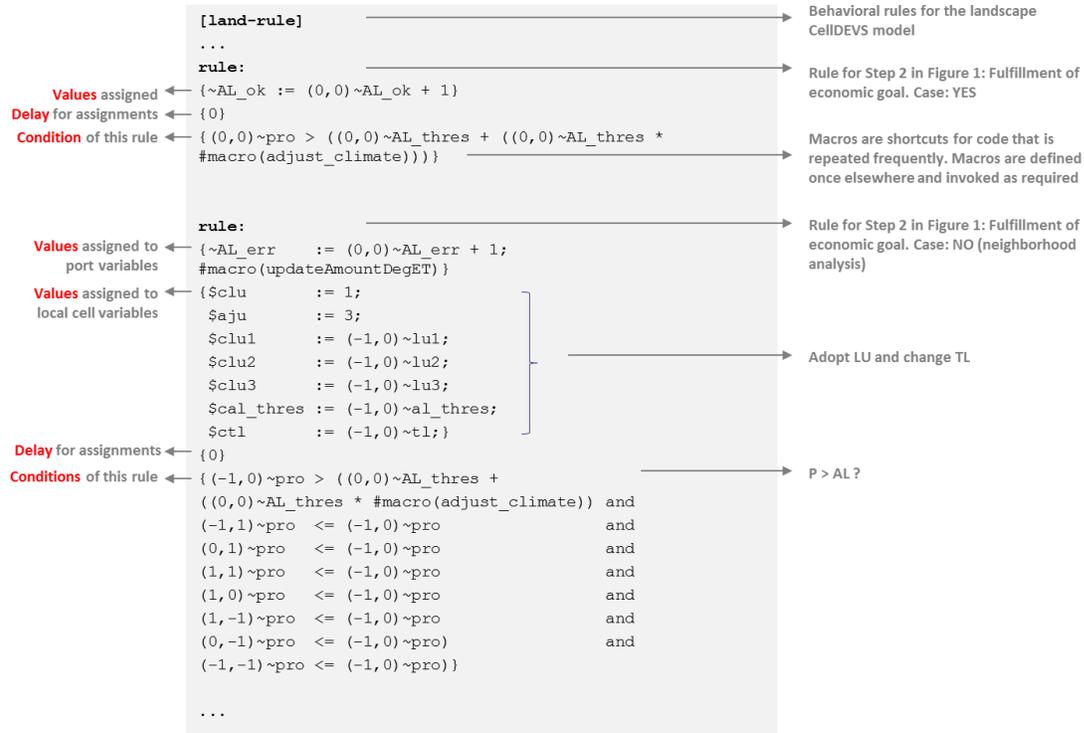

Figure E2. A sample of two rules defining the behavior of each cell of the Landscape Cell-DEVS model, showing the case for Step 2 in Figure 3 (main text).

**Simulation framework and software experimentation environment**

As discussed previously in the modelling approach section, AgroDEVS uses a DEVS-based formal approach. One salient feature of the DEVS formalism is the strict and clear separation between simulation algorithms and model specification. As a specialization of DEVS for Cellular Automata, the Cell-DEVS formalism inherits this separation. Different Cell-DEVS-capable simulators should be able to simulate a given Cell-DEVS model. In turn, different interactive user interfaces can be used to help with the design and maintenance of model specifications and to interact with the simulation exercises. Figure E3 shows a high-level component and deployment diagram of AgroDEVS.

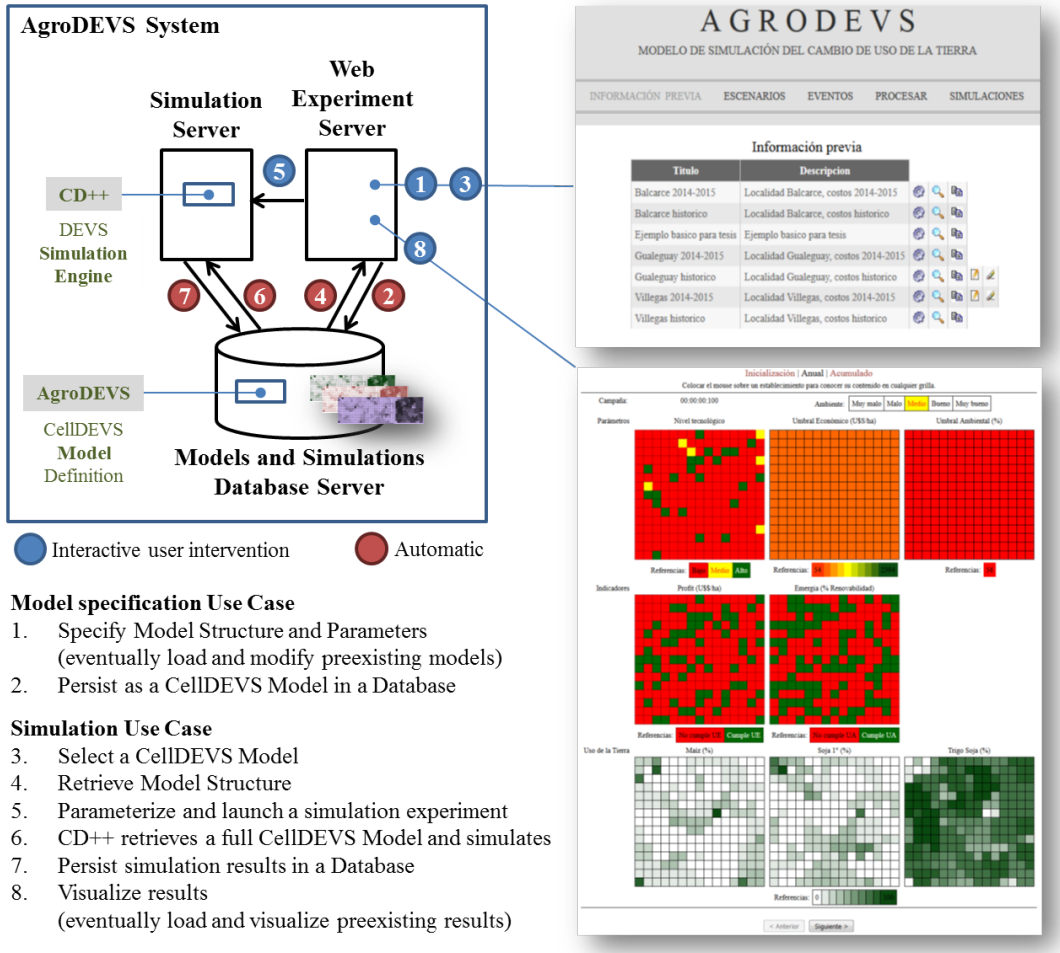

Figure E3. The AgroDEVS System. Left, top: High-level architecture and main components. Left, bottom: Typical steps and use cases. Right: Web interface for model management and results visualization

The CD++ generic simulation engine for DEVS and Cell-DEVS models resides in a Simulation Server. It can retrieve and simulate Cell-DEVS formal specifications of models that are stored in a Database Server. Such models can be written directly by specialists in the Cell-DEVS formalism or by higher-level experts in the agricultural domain using a friendly web-based interface. Once models are defined, they can be reused to launch as many simulation experiments as required, by invoking the simulator also from a web-based experiment interface. Figure E3 describes sequences of steps for two typical use cases: Model Specification and Simulation, denoting the components involved in each of them.

**Appendix F. Design concepts**

*Emergence*

Four main landscape-level attributes emerge from individual farmer's behavior and interactions among agents: (a) land use (% of cropping area under each crop type), (b) economic profit (the average landscape profit per hectare), (c) emergy renewability level (the average fraction of renewable emergy consumption in the landscape), and (d) the average rate of fulfillment of two thresholds: 1) a fixed environmental threshold, in terms of emergy renewability level (RL), and 2) a dynamic aspiration level, in terms of economic profit (P).

*Adaptation*

Agents can adjust their crop type allocation if the economic profit (at farm level) does not reach the AL at each CC. In addition, each agent has two different adaptation mechanisms: a) adopting different TL based on the capital availability, and b) adjusting its AL based on both the WGC of each CC and the outcome in the previous CC (see sub-model section for computational details).

*Objectives*

Agents pursue to achieve a P level in each CC above their AL while satisfying (or not) a fixed RL. If P is below agents' AL, they will be unsatisfied and will seek a different crop type allocation in the agents' neighborhood. If an agent's capital availability drops below the WCT for each TL, they must adopt a lower-cost TL, but agents never quit farming. Both landscape and individual RL emerge from the crop allocation decided for each agent, but it is not used as a farmer´s goal. Rather, this is an emergent property due to local rules.

*Prediction*

Agents predict the future consequences of their decisions (i.e. they build expectations) about P based on past outcomes and current weather information (WGC), following the theory of adaptive expectations (Shell and Stiglitz, 1967). Agents revise their AL in each CC using adaptive rules based on a) the dissimilarity -in the previous CC- between the predicted and the observed P, and b) the seasonal climate forecasts (WGC) for the current CC (see sub-model section for algorithms adjusting AL)

*Sensing*

Agents have information about their capital availability and consider this variable in their decisions about the potential adoption of higher (or lower) TL. In addition, they are aware of the P achieved by their eight neighbors (Moore neighborhood) during each CC. Finally, farmers are informed about the expected status of external contextual factors (i.e. seasonal forecast) for the current CC.

*Agent–Agent interaction*

Agent interaction is based on the imitation of the crop type allocation of the best neighboring agent (i.e. the highest P achieved in the neighborhood) in the case of dissatisfaction with its own P achieved (as compared to its AL). Otherwise, the agent repeats its current crop type allocation in the next CC.

*Agent–environment interaction*

Crop yield simulations reflect the interaction between farm decisions and both climate and soil conditions. These models capture the TL effect (through increasing fertilization and pesticide use) as well as the WGC of each CC (mainly global radiation and precipitation) and land quality factors (which are constant during the simulation). As crop simulation models represented biotic-unconstrained yield potential, final yields are adjusted in each combination of crop type-TL using potential weeds and pest infestation losses (see simulations input section for an example of crop yield data). Agent decisions have no effect on resource dynamics, through crop type allocation, and it is only on the current WGC that agents receive feedback about changes in resource conditions.

*Stochasticity*

Stochasticity is used to assign the TL and the crop type allocation to each agent only at the initialization procedure.

*Collectives*

Agents do not form or belong to aggregations that affect or are affected by other agents in the model version described here. Further model refinements should include collectives.